\documentclass[twoside,twocolumn,english,aps,prl,showpacs,superscriptaddress,longbibliography]{revtex4-2}
\usepackage[T1]{fontenc}

\usepackage{xcolor}
\usepackage{bm}
\usepackage{amsmath}
\usepackage{amssymb}
\usepackage{graphicx}
\graphicspath{{./}{figs/}{images/}}
\DeclareGraphicsExtensions{.pdf,.png,.jpg,.jpeg}

\usepackage{esint}
\usepackage{mathrsfs}
\usepackage{booktabs}
\usepackage{dcolumn}
\usepackage{tabularx}
\newcolumntype{Y}{>{\centering\arraybackslash}X}
\newcolumntype{b}{>{\hsize=1.2\hsize}X}

\usepackage{multirow}
\usepackage[normalem]{ulem}
\usepackage{babel}
\makeatletter
\newcommand*{\rom}[1]{\expandafter\@slowromancap\romannumeral #1@}
\makeatother

\usepackage[colorlinks=true,allcolors=blue]{hyperref}

\usepackage{balance}

\begin{document}

\title{
Conductance Anomaly in a Partially Open Adiabatic Quantum Point Contact}
\author{Donghao Liu}
\email{donghaoliu@gbu.edu.cn}
\affiliation{School of Science, Great Bay University, Dongguan, 523000, China}

\author{Dmitri Gutman}
\affiliation{Department of Physics, Bar-Ilan University, Ramat Gan, 52900, Israel}

\date{\today}

\begin{abstract}
We demonstrate that conductance anomalies can arise in a clean, adiabatic quantum point contact when a channel is partially transmitting. Even for a smooth barrier potential, backscattering induces Friedel oscillations that, via electron interactions, generate a singular correction to the conductance.  This correction is maximized when the channel is half-open, resulting in a reduction of conductance. 
In addition, a magnetic field applied perpendicular to the spin-orbit axis modifies the single-particle spectrum, resulting in conductance oscillations via Fabry-P\'erot-type interference, as well as a non-monotonic field dependence of the anomaly.
Our findings reveal a universal mechanism by which interactions modify the conductance of an ideal partially open channel and offer a possible explanation for the anomalous features observed in experiments.
\end{abstract}

\maketitle
\section{Introduction}

The quantization of Landauer conductance in units of universal quantum conductance  $g_0\equiv e^{2}/h$ is a cornerstone of mesoscopic transport. For fully open channels, this quantization persists even in the presence of electron-electron interactions.
Galilean invariance  guarantees that  the total momentum of electrons is proportional to the current 
${\bf J}=e/m_e {\bf P}$. Thus the conservation of the total momentum implies the conservation of the electric current, hence conductance. 
This universality was first demonstrated for the  Landauer-Sharvin conductance~\cite{levinson1989potential} within the quantum-kinetic approach. 
This result was later generalized for the Luttinger liquid geometry ~\cite{maslov1995landauer,ponomarenko1995renormalization,safi1995transport,oreg1995interedge,ponomarenko1996frequency,safi1999interacting,trauzettel2004appearance}. 

 By contrast, in the presence of a weak link or a localized impurity, which introduces a finite bare backscattering amplitude and thus renders the channel only partially transmitting, electron-electron interactions can profoundly modify transport, as demonstrated in Refs.~\cite{kane1992transport,kane1992transmission}; for review see, e.g., Ref.~\cite{MaslovLH2005}.
Throughout this work, we
characterize the degree of channel opening by the transmission
probability $\mathcal{T}$: a channel is referred to as \emph{open} when
$\mathcal{T} \simeq 1$, as \emph{closed} when $\mathcal{T} \simeq 0$, as
\emph{partially open} when $0 < \mathcal{T} < 1$, and as \emph{half-open}
when $\mathcal{T} \sim 0.5$.

Short junctions can be modeled as effective impurities in the infrared (low-energy) limit; under renormalization, they flow toward either the pinch-off (perfect reflection) fixed point or the perfectly transmitting fixed point, as described in Refs.~\cite{kane1992transport,kane1992transmission}. In our terminology, these correspond to the closed- and open-channel limits, respectively.
In such systems, interaction effects originate from the renormalization of
a pre-existing scattering potential. By contrast, long and adiabatic channels are commonly assumed to exhibit negligible backscattering, so one would not expect pronounced interaction fingerprints if the channel is fully  opened. Nevertheless, a number of experiments in this latter regime report a striking, nontrivial evolution of the conductance as the channel opens.
These unexpected plateaus appear at simple fractions of the conductance quantum,
 even when the channel is clean and adiabatic~\cite{kumar2019zero}.  Similar fractional plateaus have since been observed in related one-dimensional platforms, including hole quantum wires~\cite{gul2018self} and high-mobility InGaAs heterostructures~\cite{liu2023possible,rodriguez2025nonmagnetic}, indicating that the phenomenon is not limited to a single material system or growth technique.  

To account for these anomalies, a variety of mechanisms have been proposed.  These include the formation of a quasi-localized state inside the constriction, producing a Kondo-like zero-bias peak and scaling behavior~\cite{Cronenwett2002,Meir2002,rejec_magnetic_2006,iqbal_odd_2013}; interaction-enhanced spin splitting or a spin gap~\cite{Reilly2005,Hudson2021}; and a van-Hove-type density-of-states ridge at the barrier top that amplifies spin fluctuations and unifies the fractional plateaus with the conventional $0.7 \times 2g_0$   shoulder and zero-bias anomaly~\cite{Bauer2013}.  

Device electrostatics strongly tune the strength and position of these features~\cite{Smith2015}. In particular, the plateau value decreases from $0.7$ to $0.6$ as the effective barrier length increases, suggesting that there is no universality in the number itself. 
Shot-noise measurements~\cite{DiCarlo2006} reveal a suppression of partition noise, in contrast to the Landauer-B\"uttiker-Lesovik prediction~\cite{lesovik1989excess}, which, together with the systematic magnetic-field dependence, highlights the important role of the spin degree of freedom.
 While these results pertain to moderately interacting electrons, strong interactions may drive the system toward a charge-density-wave ground state, also known as a Wigner crystal. In this regime, spin-incoherent Luttinger liquids predict a suppression of ideal quantization and shifts of plateau values~\cite{Matveev2004PRL,Matveev2004PRB,FieteRMP2007}. Moreover, conductance fractionalization can occur in strongly interacting Luttinger liquids, where multi-particle backscattering processes are relevant~\cite{shavit2019fractional,shavit2020electron,shavit2020modulation}.

Despite this impressive progress in understanding,  it remains of fundamental interest to ask what universal features of conductance can be expected for interacting electrons in an ideal Landauer channel, i.e., adiabatic, yet not fully open.
In this work, we address precisely this regime. We consider a clean, adiabatic Landauer contact and analyze how interactions alter its conductance as the channel opens.

We find that the correction to the conductance arises from the interplay of two effects:

(i) Electron backscattering can occur even for a smooth potential barrier and at energies above the barrier, leading to Friedel oscillations in the electron density.

(ii) These oscillations, characterized by a wave vector $2k_F$, resonantly enhance backscattering. In effect, they generate a spatially modulated potential with a $2k_F$ component, thereby invalidating the smooth-barrier approximation and producing a singular correction to the conductance.

Our analysis suggests that this mechanism can account for salient features observed in recent experiments.
We further show that the correction persists when either a magnetic field or spin-orbit coupling (SOC) dominates, but is strongly suppressed when the two coexist with mutually perpendicular components.

\section{Model and scattering states}
We consider an adiabatic point contact described by the effective one-dimensional Hamiltonian
$H=\sum_{i}h(x_i)+\frac{1}{2}\sum_{i\neq j} U(x_i-x_j)$, where $h(x_i)$ is the single-particle Hamiltonian and $U(x_i-x_j)$ denotes the electron-electron interaction.  
The single-particle part reads
\begin{equation}
h(x)=\left[-\frac{\hat{p}^2}{2m_e^{*}}+V(x)-\mu\right]\sigma_0
+ \hat{p}\,\boldsymbol{\gamma}\cdot\boldsymbol{\sigma}
+ g\mu_B \boldsymbol{B}\cdot\boldsymbol{\sigma},\label{eq:H0}
\end{equation}
with momentum operator $\hat{p}=-i\hbar d/dx$, Pauli matrices $\sigma_{0,x,y,z}$, effective mass $m_e^*$, and chemical potential $\mu$.  
The second term represents spin-orbit coupling, while the last term is the Zeeman coupling to the magnetic field $\boldsymbol{B}$, with $g$ the Land\'e factor and $\mu_B$ the Bohr magneton.  

We emphasize that the inclusion of SOC and magnetic field allows us to study more general scenarios.  
However, the central conclusions of this work, in particular the interaction-induced Friedel oscillations and the resulting conductance anomalies, remain valid even in the absence of SOC.  
For concreteness, we model the barrier potential $V(x)$ as a P\"{o}schl-Teller form,
\begin{equation}
V(x)=-\frac{\hbar^2}{2m_e^*}\frac{\alpha^2\lambda(\lambda-1)}{\cosh^2(\alpha x)}.\label{eq:Vx}
\end{equation}
The corresponding profile is shown in Fig.~\ref{fig:Density}(a); only the $x<0$ side is plotted since $V(x)$ is symmetric about $x=0$.
Here, $\lambda=1/2+il$ and $\alpha$ control the barrier height $V(0)=\frac{\hbar^2}{2m_e^{*}}\alpha^2(l^2+\frac{1}{4})$
and curvature 
$V^{\prime\prime}(0)
=-\frac{\hbar^2}{m_e^{*}}\alpha^4\left(l^2+\frac{1}{4}\right)$.
These are two essential characteristics of the confining potential in the quantum point contact.
Within the semiclassical approximation (WKB), one defines an effective length 
$\ell(\epsilon)\equiv\left|2(\epsilon-V(0))/V^{\prime\prime}(0)\right|^{1/2}$.  
While the  potential in Eq.~(\ref{eq:Vx}) admits an analytic solution~\cite{CevikPLA2016}, the WKB analysis applies to the
broader class of potential barriers~\cite{BerryMount1972} with the main features unchanged.
Here, $\epsilon$ is the energy of the scattering states measured from the band bottom.
 
Far away from the constriction, the system becomes translationally invariant on both sides. We denote the corresponding asymptotic regions as the left (L) and right (R) sides. 
The scattering states $\Psi^L_s$ and $\Psi^R_s$ are solutions of the Hamiltonian in Eq.~(\ref{eq:H0}), where the superscript $L$ ($R$) specifies the side of incidence and  the subscript $s=1,2$ is a channel index labeling the two propagating modes.
In the asymptotic regions, they take the form:
\begin{align}
\Psi_{s,k}^{L}\left(x\right)=\begin{cases}
\phi_s^+(x)+\sum_{s'}r^{L}_{s's}\phi_{s'}^-(x), & x \ll -\ell(\epsilon),
\\[10pt]
\displaystyle\sum_{s'}
t^{L}_{s's}\,\phi_{s'}^{+}(x), & x\gg \ell(\epsilon),
\end{cases}
\label{eq:PhiL}
\end{align}

\begin{align}
\Psi^{R}_{s,k}(x)
=
\begin{cases}
\displaystyle\sum_{s'}
t^{R}_{s's}\,\phi_{s'}^{-}(x), & x\ll -\ell(\epsilon),\\[10pt]
\phi_s^{-}(x)
+
\displaystyle\sum_{s'}
r^{R}_{s's}\,\phi_{s'}^{+}(x), & x\gg \ell(\epsilon).
\end{cases}
\label{eq:PhiR}
\end{align} 
The derivation is given in Supplementary Material, Sec.~\ref{SI:eigenstates}.
Here the parameter $k>0$ is used to parametrize the energy $\epsilon$ of the scattering state, with $\epsilon=\hbar^2 k^2/(2m_e^*)$.
We define
$\phi_s^{\pm}(x)=e^{ik_s^{\pm}x}\,|\chi_s^{\pm}\rangle$.
For each channel index $s=1,2$, the dispersion relation of the translationally invariant Hamiltonian~(\ref{eq:H0}) in the absence of the barrier yields two propagation wave vectors $k_s^{\pm}$ at energy $\epsilon$, corresponding to positive $(+)$ and negative $(-)$ group velocities. 
At the Fermi energy, there are four Fermi wave vectors $\{k_s^{\pm}\}$.
The spinors $|\chi_s^{\pm}\rangle$ are the corresponding eigenvectors of the translationally invariant Hamiltonian in momentum space evaluated at $k_s^{\pm}$.
The matrices $r^{L,R}$ and $t^{L,R}$ are $2\times2$ reflection and transmission matrices. 
They generally depend on the energy parameter $k$ through the dispersion relation; for brevity, this dependence is suppressed unless explicitly required. 
Mode mixing between different channels can occur due to SOC and the magnetic field.

Note that the long-distance behavior of the wave function for scattering from a smooth potential is qualitatively similar to that for a short-range potential. The main difference is encoded in the energy-dependent scale $\ell(\epsilon)$ at which the asymptotic expansion Eqs.~(\ref{eq:PhiL}) and (\ref{eq:PhiR}) can be applied.

\section{Zero-field conductance anomaly}
We first analyze the problem in the absence of a magnetic field.

\subsection{Scattering states}
In this limit, time-reversal symmetry is preserved, and the channels form Kramers pairs.
Consequently, scattering does not mix different channels, and the reflection and transmission matrices $r^{L,R}$ and $t^{L,R}$ become diagonal, $r^{L,R}_{s's}=r^{L,R}_{s}\delta_{s's}$, $t^{L,R}_{s's}=t^{L,R}_{s}\delta_{s's}$.
Accordingly, each scattering channel 
$s$ corresponds to a well-defined spin state. The eigenvectors satisfy $|\chi_s^{+}\rangle=|\chi_s^{-}\rangle$ and are eigenvectors of $\boldsymbol{\gamma}\cdot\boldsymbol{\sigma}$, independent of momentum.
The wave vectors $k_{1,2}^{\pm}$ are 
 $k_1^{\pm}=\pm k- k_{\gamma}$ and  $ k_2^{\pm}=\pm k+ k_{\gamma}$, where $k_{\gamma}=m_e^{*} \gamma/\hbar$ and $k=\sqrt{2m_e^*\epsilon}/\hbar>0$.

In this situation, the scattering potential is spin independent, and the two helicity channels are therefore equivalent, resulting in identical reflection and transmission amplitudes\cite{CevikPLA2016} for $s=1,2$:
\begin{equation}
r^{L,R}_{s}(\tilde{k})=\frac{\Gamma\left(i\tilde{k}\right)\Gamma\left(1-i\tilde{k}-\lambda\right)\Gamma\left(-i\tilde{k}+\lambda\right)}{\Gamma\left(\lambda\right)\Gamma\left(1-\lambda\right)\Gamma\left(-i\tilde{k}\right)}, \label{eq:rL}
\end{equation}
\begin{equation}
t^{L,R}_{s}(\tilde{k})=\frac{\Gamma\left(1-i\tilde{k}-\lambda\right)\Gamma\left(-i\tilde{k}+\lambda\right)}{\Gamma\left(-i\tilde{k}\right)\Gamma\left(1-i\tilde{k}\right)}, \label{eq:tL}
\end{equation}
where $\Gamma\left(z\right)$ is the gamma function, and $\tilde{k}={k}/\alpha$. The corresponding transmission probability is $\mathcal{T}_{s}(\tilde{k})=\left|t^L_{s}(\tilde{k})\right|^2$.

The pairs $(k_1^{+},k_2^{-})$ and $(k_2^{+},k_1^{-})$ are Kramers partners. 
Time-reversal symmetry forbids scattering between such partners; hence, in the absence of time-reversal breaking, backscattering occurs only within a given mode, $k_1^{+}\leftrightarrow k_1^{-}$ or $k_2^{+}\leftrightarrow k_2^{-}$, with no mixing between modes 1 and 2.
This is precisely the structure encoded in Eqs.~(\ref{eq:PhiL})--(\ref{eq:PhiR}).
When a magnetic field is applied, it breaks the time-reversal symmetry, enabling  $k_1\leftrightarrow k_2$ scattering and promoting the reflection and transmission amplitudes to matrices.
 

\begin{figure}[t]
 \centering
\includegraphics[width=1\columnwidth]{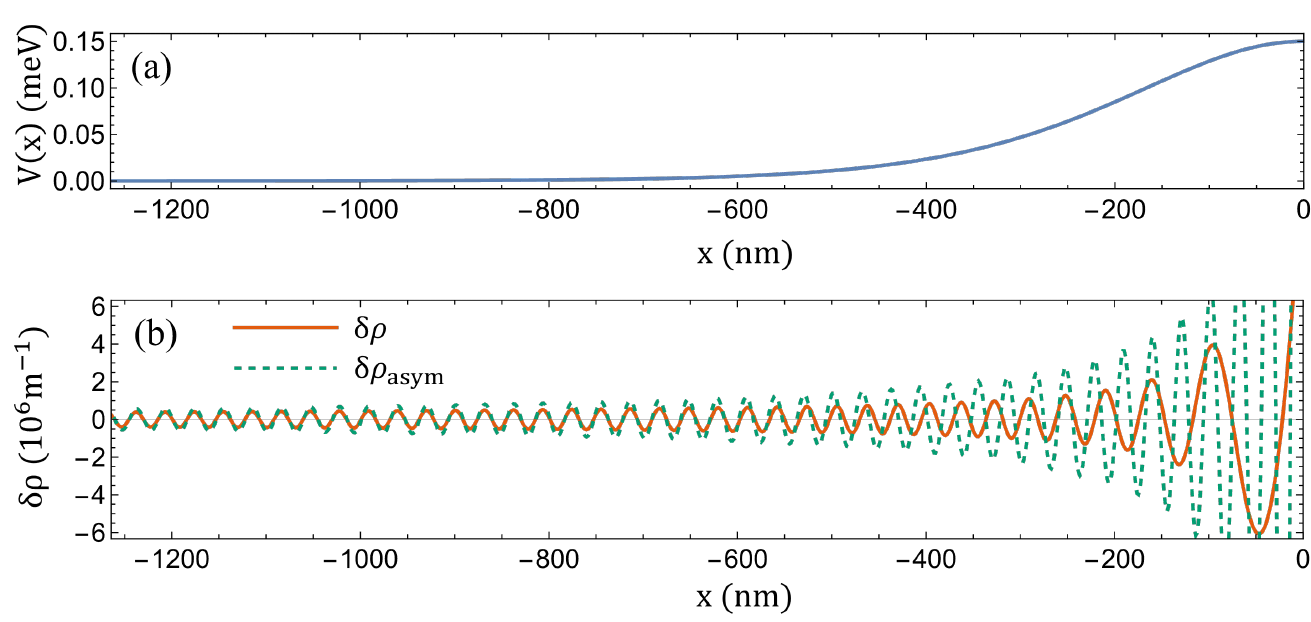}
\caption{
(a) The barrier potential $V\left(x\right)$ with the model parameter set by $m_e^{*}=0.067m_e$ (with $m_e$ the electron mass),  $\lambda=1/2+il$, $l=25.8$, and $\alpha=3.95\times 10^{-3}\,\mathrm{nm}^{-1}$.
Only the $x<0$ region is shown, since the potential 
$V(x)$ is symmetric about 
$x=0$.
Panel (b) shows the corresponding density as a function of $x$. 
$\delta\rho$ is obtained from the exact analytic solution of the Schrödinger equation for the Hamiltonian in Eq.~(\ref{eq:H0}), in contrast to the asymptotic result
$\delta\rho_{asym}$ in Eq.~(\ref{densityosc}).}
\label{fig:Density}
\end{figure}

\subsection{Friedel oscillations }
Scattering modifies the local electron density $\rho\left(x\right)=\sum_{s,k<k_F}\Psi_{s,k}^{R\dagger}\Psi_{s,k}^{R} + \Psi_{s,k}^{L\dagger}\Psi_{s,k}^{L}$, which can be decomposed as:  $\rho\left(x\right) = \bar{\rho}\left(x\right)+\delta\rho\left(x\right)$. 
Here, $\bar{\rho}(x)$ is the smooth background density, while $\delta\rho(x)$ denotes the reflection-induced oscillatory correction (Friedel oscillations). 
Far from the barrier, for $|x|\gg \ell(\epsilon)$, $\delta\rho(x)$ approaches its asymptotic form.
\begin{equation}
\delta\rho_{\mathrm{asym}}\left(x\right)=
 \frac{\left|r_{F}\right|}{\pi x}
\begin{cases}
\sin{\left(2k_F x-\theta_{F}\right)}, & x\ll -\ell(\epsilon),\\[8pt]
\sin{\left(2k_F x+\theta_{F}\right)} , & x\gg \ell(\epsilon).
\end{cases}
\label{densityosc}
\end{equation}
Here,
$k_F=\sqrt{2m_e^*\mu}/\hbar$, $r_F\equiv r_s^{L}(k_F)=r_s^{R}(k_F)$ and $\theta_F\equiv \arg r_F$.
The asymptotic form $\delta\rho_{\mathrm{asym}}(x)$ from Eq.~(\ref{densityosc})
is compared with the exact result $\delta\rho(x)$ obtained from the analytic scattering states
of the Hamiltonian~(\ref{eq:H0}) in Fig.~\ref{fig:Density}(b).
They coincide far from the barrier, $|x|\gg \ell(\epsilon)$, and differ only near it,
$|x|\lesssim \ell(\epsilon)$, where the asymptotic approximation breaks down.
As we will show below, the long-distance Friedel-oscillation tail $\delta\rho(x)\propto 1/|x|$ is what gives rise to the logarithmic conductance correction, so short-distance deviations near the barrier do not modify the leading log term.

\subsection{Interaction-induced conductance correction}
We now compute the interaction correction to the conductance induced by the Friedel oscillations, following the approach of Ref.~\cite{MatveevYueGlazman1993PRL} developed for short-range scattering potentials, and adapt it to the present smooth barrier.


The electron-electron interaction $U\left(x-y\right)$  produces two principal effects. 
First, it screens the static barrier $V\left(x\right)$, yielding a smoother effective profile, which can still be modeled by Eq.~(\ref{eq:Vx}). Since the bare barrier is not directly observable, it is natural to treat the parameters of the effective P\"oschl-Teller potential as already renormalized by self-consistent screening.

Second, and more importantly, the Friedel oscillations of density $\delta\rho\left(x\right)$ 
generate an additional $2k_F$ component of the potential. Starting from the two-body interaction term $U(x_i-x_j)$ in the
many-body Hamiltonian, we express the interaction in terms of the
electron density and adopt a continuum description.
Within the Hartree approximation, this contribution is given by $V_H\left(y\right)=\int_{-\infty}^{\infty}U\left(y-z\right)\delta\rho\left(z\right)dz$. In the following, we approximate the interaction as short-ranged, $U(x-y)\simeq\beta\pi\hbar v_{F}\delta(x-y)$, with $\beta$ the dimensionless interaction strength.
Being resonant with the Fermi momentum, the oscillating potential $V_H$ induces a singular correction to the conductance.

The correction to the wave function induced by the oscillating potential can be obtained within the Born approximation,
\begin{equation}
\delta\Psi_{s,k}^{L}\left(x\right)=\int_{-\infty}^{\infty}G_{s,k}^{L}\left(x,y\right)V_H\left(y\right)\Psi_{s,k}^{L}\left(y\right)dy, \label{eq:deltaPhi}
\end{equation}
where $G_{s,k}^L$ is the Green's function of noninteracting electrons in the presence of the barrier potential:
\begin{equation}
G_{s,k}^{L}\left(x,y\right)=\frac{e^{ik_{s}^{-}x}}{i\hbar v_{k}}\begin{cases}
e^{-ik_{s}^{-}y}+r^L_{s}e^{-ik_{s}^{+}y}, & y\ll-\ell(\epsilon),\\
t^R_{s} e^{-ik_{s}^{-}y}, & y\gg \ell(\epsilon).
\end{cases}
\nonumber
\end{equation}
Here $v_k$ is the electron velocity.  
Substituting $G_{s,k}^{L}$ and $V_H$ into Eq.~(\ref{eq:deltaPhi}) yields the correction to the scattering state:
$\delta\Psi_{s,k}^{L}(x)=\delta r_s^L \phi_s^-(x)$,
which has the structure of the reflected wave and defines the interaction-induced renormalization $\delta r_s^L$ of the reflection amplitude:
\begin{equation}
\delta r_s^L=\beta\mathcal{T}_s r_s^L \ln\left({\frac{1}{\left|k-k_F\right|L}}\right). \label{eq:deltar}
\end{equation}

This enhanced reflection amplitude  leads to a reduction of the transmission:
\begin{equation}
\delta \mathcal{T}_{s}(k)=-2\beta\mathcal{T}_{s}(k)\left[1-\mathcal{T}_{s}(k)\right]\ln\left({\frac{1}{\left|k-k_F\right|L}}\right).
\label{eq:delta_T}
\end{equation}
The ultraviolet cutoff length $L$ is set by the largest of the characteristic spatial scales:    
the effective width of the barrier  at the Fermi energy  $\ell(\epsilon_F)$, the Fermi wavelength far from the barrier 
$\lambda_F$,  and the characteristic scale of the short-range interaction.

The correction in Eq.~(\ref{eq:delta_T}) produces an anomaly in the conductance. 
The prefactor $\mathcal{T}_{s}(1-\mathcal{T}_{s})$ in Eq.~(\ref{eq:delta_T}) implies that the anomaly reaches its maximum at $\mathcal{T}_{s}=1/2$, i.e., when the chemical potential is aligned with the barrier top and the channel is half-open.  
Physically, this dependence reflects that the interaction correction requires both a finite reflection (to generate Friedel oscillations in the density) and a finite transmission (so that the additional backscattering modifies transport). Hence the correction vanishes deep on the plateaus, $\mathcal{T}_s\to 0$ or $1$, and is strongest in the transition regime where the channel is partially transmitting.
At finite temperature, the conductance correction is
\begin{equation}
\delta G = \frac{e^2}{h}\int d\epsilon \sum_{s}\delta \mathcal{T}_s(\epsilon)\left( -\frac{\partial f_{\epsilon}}{\partial \epsilon} \right),
\end{equation}
where $f_{\epsilon}$ is the Fermi-Dirac distribution.
The thermal kernel $-\partial f_\epsilon/\partial\epsilon$ restricts the integral to $|\epsilon-\mu|\sim k_BT$, thereby providing the infrared cutoff of the logarithm without introducing an additional phenomenological scale.
Here $\delta G$ denotes the interaction-induced correction to the conductance. The corresponding total conductance is $G(\mu)=G_0(\mu)+\delta G(\mu)$, where $G_0$ is the noninteracting conductance. The magnitude of $\delta G$ determines how prominently the anomaly appears in the total conductance line shape.

\begin{figure}[t]
 \centering
\includegraphics[width=1\columnwidth]{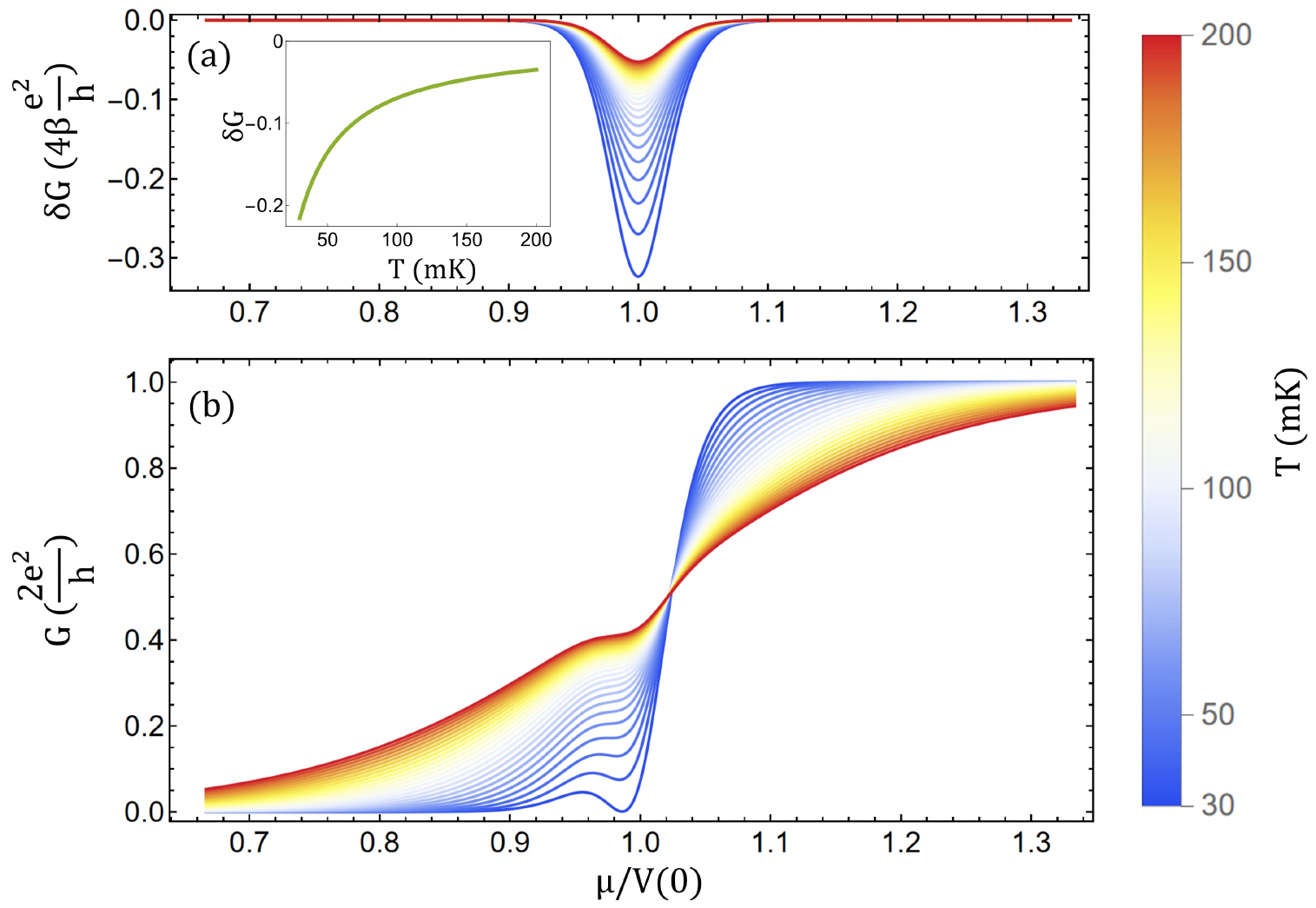}
\caption{
Panel (a) shows the anomalous conductance $\delta G$ and panel (b) shows the total conductance $G=G_0+\delta G$, both as functions of chemical potential for several temperatures. Inset of (a): minimum $\delta G$ versus temperature. $V(0)$ denotes the barrier maximum in Eq.~(\ref{eq:Vx}). Barrier parameters are the same as in Fig.~\ref{fig:Density}. Here we take the dimensionless interaction strength $\beta=1$.}
\label{fig:AnomalyVsChemicalPotential}
\end{figure}

Fig.~\ref{fig:AnomalyVsChemicalPotential}(a) shows $\delta G$ as a function of chemical potential for several temperatures. The anomalous correction is expressed in units of $4\beta e^2/h$. The factor of 4 arises because, in the zero-field case, Friedel oscillations generated by either spin channel resonantly backscatter electrons in both channels. Since the two channels are equivalent, these four contributions are equal. As the temperature is lowered, $|\delta G|$ increases, reflecting the reduced infrared cutoff and the corresponding enhancement of the interaction-induced anomaly. 
Fig.~\ref{fig:AnomalyVsChemicalPotential}(b) shows the total conductance $G(\mu)$ versus chemical potential at the same temperatures. Between two conductance plateaus, $\delta G$ produces a pronounced suppression of the conductance in the transition region, where the anomaly is most visible.

While the existence and functional form of the anomaly follow from a universal mechanism---interaction-induced Friedel oscillations generate enhanced backscattering, leading to a correction controlled by the characteristic $\mathcal{T}(1-\mathcal{T})$ dependence and strengthened as the infrared cutoff is reduced---its absolute magnitude is not universal. Within our leading-log treatment, the overall size of $\delta G$ scales with the interaction strength (e.g., $\beta$) and depends on the ultraviolet and infrared cutoffs entering the logarithm, $\ln\!\left[1/\bigl(|k-k_F|\,L\bigr)\right]$. 
The ultraviolet cutoff is set by the barrier width; we take $L=500\,\mathrm{nm}$ as a representative value corresponding to typical experimental conditions, while the infrared cutoff is provided by the finite-$T$ convolution above.
Accordingly, the universal content is the mechanism and its functional dependence, whereas the overall prefactor and the detailed crossover scales are expected to be sample dependent.

\begin{figure}[t]
 \centering
\includegraphics[width=1\columnwidth]{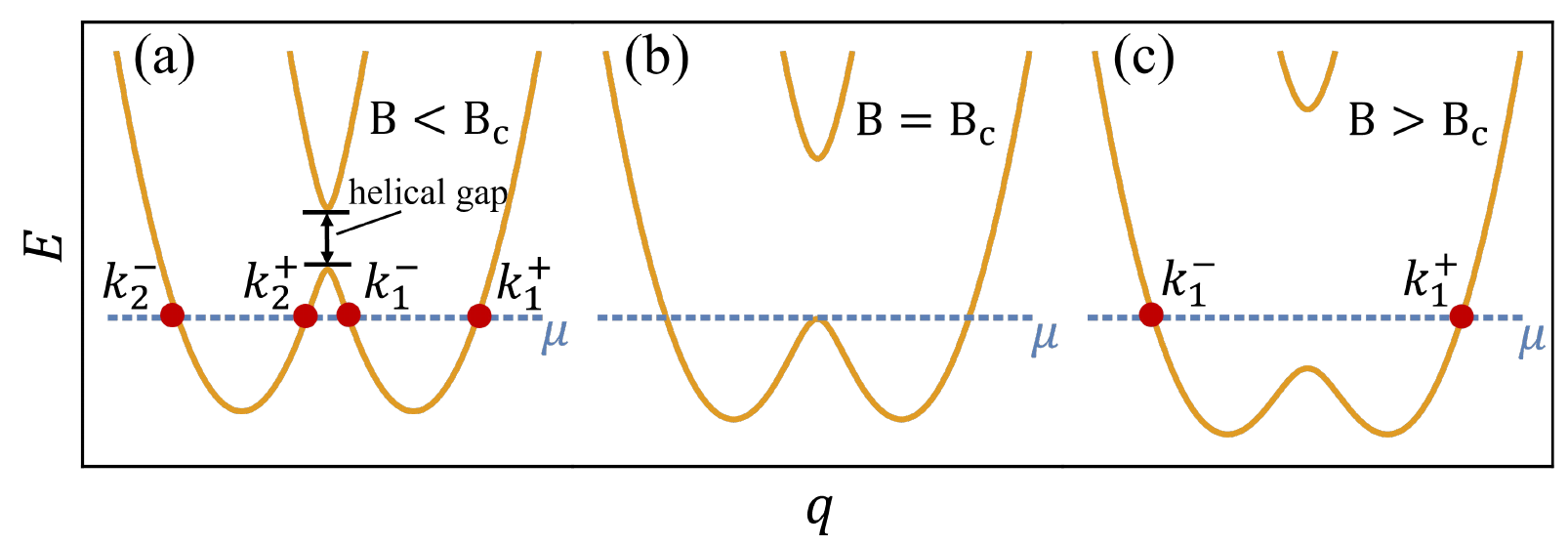}
\caption{
Band dispersion in the asymptotic region far from the central scattering region for $\boldsymbol{B}\perp\boldsymbol{\gamma}$, shown for three representative magnetic-field strengths $B<B_c$, $B=B_c$, and $B>B_c$. 
Panels (a) and (c) correspond to the weak- and strong-field regimes, respectively, while (b) illustrates the critical case $B=B_c$ where the chemical potential $\mu$ touches the hump of the lower band. 
In (a) and (c), the Fermi points that contribute to the Friedel oscillations are marked.
\label{fig:Dispersion}
}
\end{figure}

\section{Magnetic-field effects}
Next, we consider the effect of a finite magnetic field, starting with a field parallel to the spin--orbit interaction.

\subsection{Parallel field: $\boldsymbol{B}\parallel\boldsymbol{\gamma}$}
\begin{figure}[t]
 \centering
\includegraphics[width=1\columnwidth]{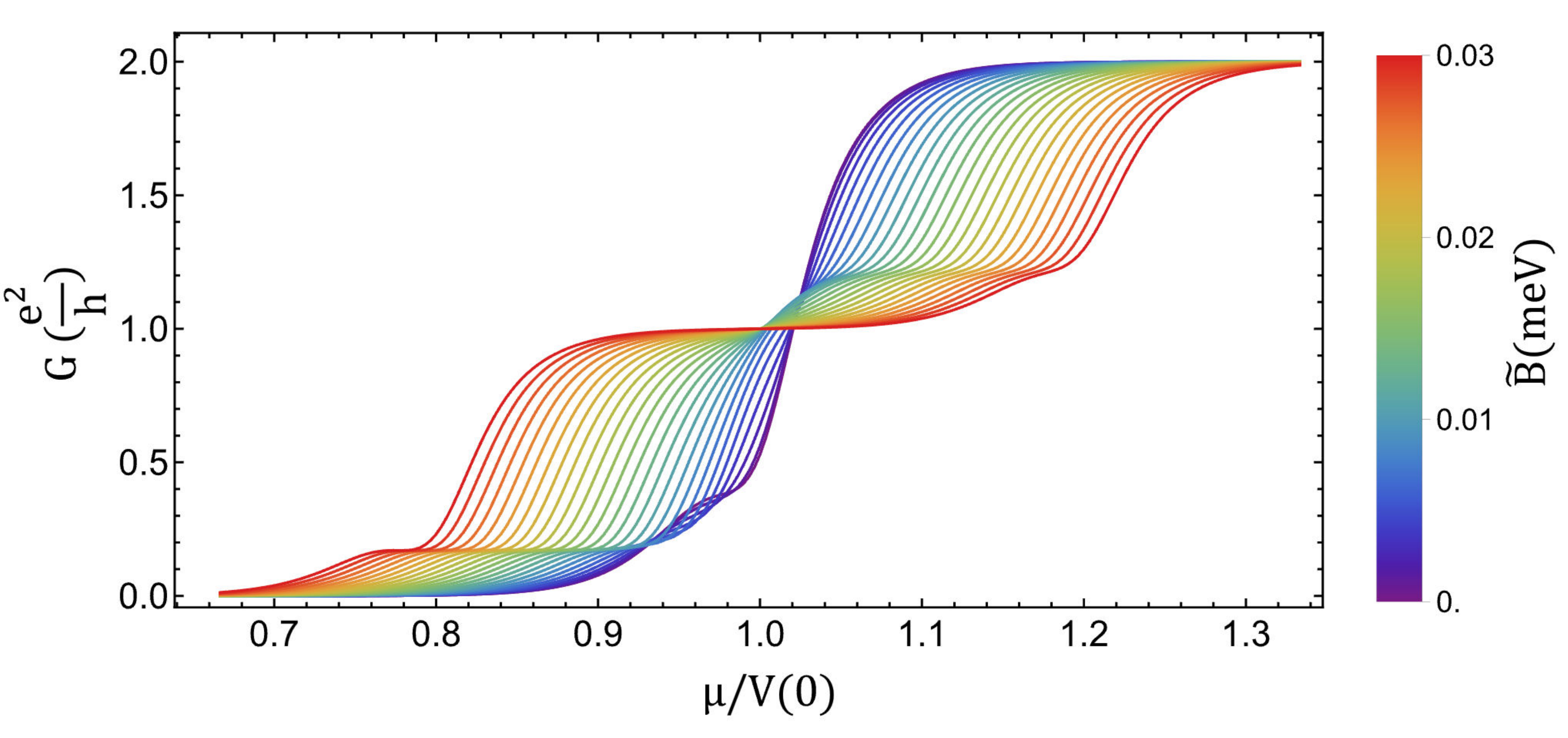}
\caption{
Total conductance (including the logarithmic correction) as a function of chemical potential for several magnetic-field strengths applied parallel to the spin–orbit-coupling axis ($\boldsymbol{B}\parallel\boldsymbol{\gamma}$).
Here $\tilde{B}=g\mu_B B$ and $T=50{\rm mK}$. 
 \label{fig:ParallelB}
}
\end{figure}

A field component parallel to the SOC axis does not affect the spin polarization and only lifts the Kramers degeneracy.
Consequently, the quantized conductance plateau at $2g_0$ splits into two successive plateaus separated by a step of $g_0$; the transitions between these two plateaus occur at distinct chemical potentials. At each transition there is a corresponding anomalous conductance.  
The total conductance $G(\mu)$, including the interaction-induced correction, is plotted in Fig.~\ref{fig:ParallelB} for several values of $B$ in the $\boldsymbol{B}\parallel\boldsymbol{\gamma}$ configuration, illustrating the field-induced splitting of the anomaly into two transition regions.

\subsection{Perpendicular field: $\boldsymbol{B}\perp\boldsymbol{\gamma}$}
We now discuss the case of a magnetic field perpendicular to the SOC axis, starting with the single-particle spectrum:
\begin{equation}
E_{\pm}(q)={\hbar^2 q^2}/{(2m_e^*)}\pm\sqrt{\hbar^2\gamma^2q^2+g^2{\mu_B}^2B^2},
\end{equation}
which is schematically shown in Fig.~\ref{fig:Dispersion}.
When $B\neq 0$, a helical gap opens at $q=0$. 
As the magnetic field increases, the lower band shifts downward, and the upper band shifts upward. 

When the lower band minimum at the top of the barrier crosses the chemical potential,  the conductance $G_0$ rises to the next plateau.
The number of Fermi points in the asymptotic region is controlled by the scale 
$g\mu_B B_c = m_e^*\gamma^2-\sqrt{2m_e^*V(0)}\,\gamma$. At $B>B_c$, $G_0$ changes in steps of $g_0$ (see Fig.~\ref{fig:NoninteractingG}).

 More significantly, the  perpendicular field leads to richer phenomena:
(i) the conductance oscillates with magnetic field even at the noninteracting level; (ii) the spin polarization at the Fermi points depends on the magnetic field.
This dependence affects the spin overlap between Fermi points, resulting in a non-monotonic dependence of the anomalous conductance on $B$.
For simplicity, in the following we focus on the strictly perpendicular case.
In this case,  the channel index $s$ is no longer conserved and scattering induces mixing between the two channels.  
Accordingly, the reflection and transmission coefficients recover the full $2\times2$ matrix structure given in Eqs.~(\ref{eq:PhiL}) and (\ref{eq:PhiR}), and the scattering problem no longer admits a closed-form solution. We therefore treat it numerically and briefly outline the procedure.

\subsubsection{Computational approach}
We now compute  the full energy- and field-dependent scattering matrix $S(\epsilon,B)$ numerically using a tight-binding discretization of the continuum Hamiltonian and the quantum-transport package \textsc{Kwant}~\cite{groth2014kwant}. 
The discretized model implements the same continuum model in Eq.~(\ref{eq:H0}) used throughout the paper; the asymptotic regions are taken uniform so that incoming/outgoing modes are well defined. 
From $S(\epsilon,B)$ we extract the transmission and reflection matrices $t_{ss'}(\epsilon,B)$ and $r_{ss'}(\epsilon,B)$ and compute the noninteracting conductance via the Landauer formula. 
We use the numerically computed scattering matrix in two ways: first, to analyze the noninteracting Landauer conductance; second, as input to the analytic expressions for the interaction-induced correction $\delta G$ derived in the Supplementary Material, Sec.~\ref{SIsec:channelmixing}.
We now turn to the Landauer conductance. Even this simple problem becomes nontrivial when spin--orbit coupling and a magnetic field are both present.

\subsubsection{Noninteracting magnetoconductance: Fabry--P\'erot-like interference}

\begin{figure}[t]
 \centering
\includegraphics[width=1\columnwidth]{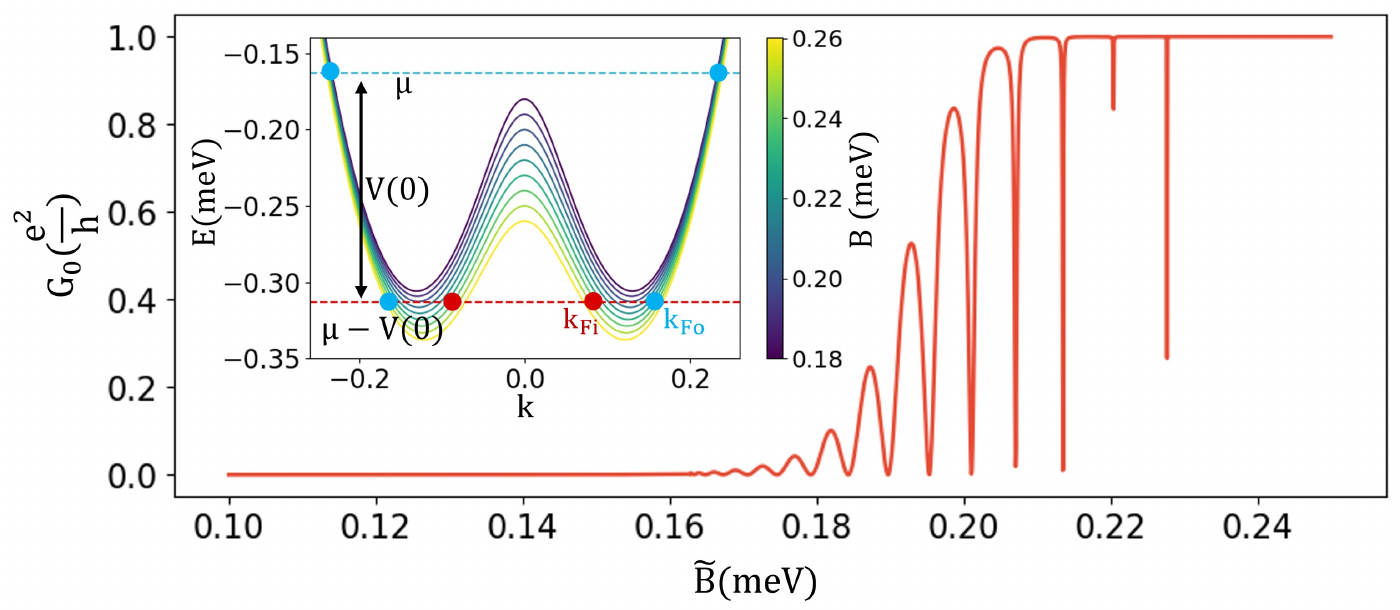}
\caption{
Noninteracting conductance $G_0$ as a function of $B$. 
 $\boldsymbol{B}\perp\boldsymbol{\gamma}$ and $\hbar\gamma=4\mathrm{meV\cdot nm}$.
As $B$ increases, the first channel opens and $G_0$ exhibits oscillations.
The chemical potential is a constant, and other parameters are the same as in Fig.~\ref{fig:AnomalyVsChemicalPotential}. 
Inset: The single-particle spectrum.  
The blue dashed line marks the chemical potential. 
Far from the barrier, it intersects the bands at two Fermi points, corresponding to a single transport channel. 
At the barrier top $V\left(0\right)$, the red dashed line marks the kinetic energy $\mu-V\left(0\right)$, which defines four local Fermi points. 
This difference in the local number of Fermi points underlies the Fabry-P\'erot-like interference discussed in the text.
\label{fig:NoninteractingG}}
\end{figure}

\begin{figure}[t]
 \centering
\includegraphics[width=1\columnwidth]{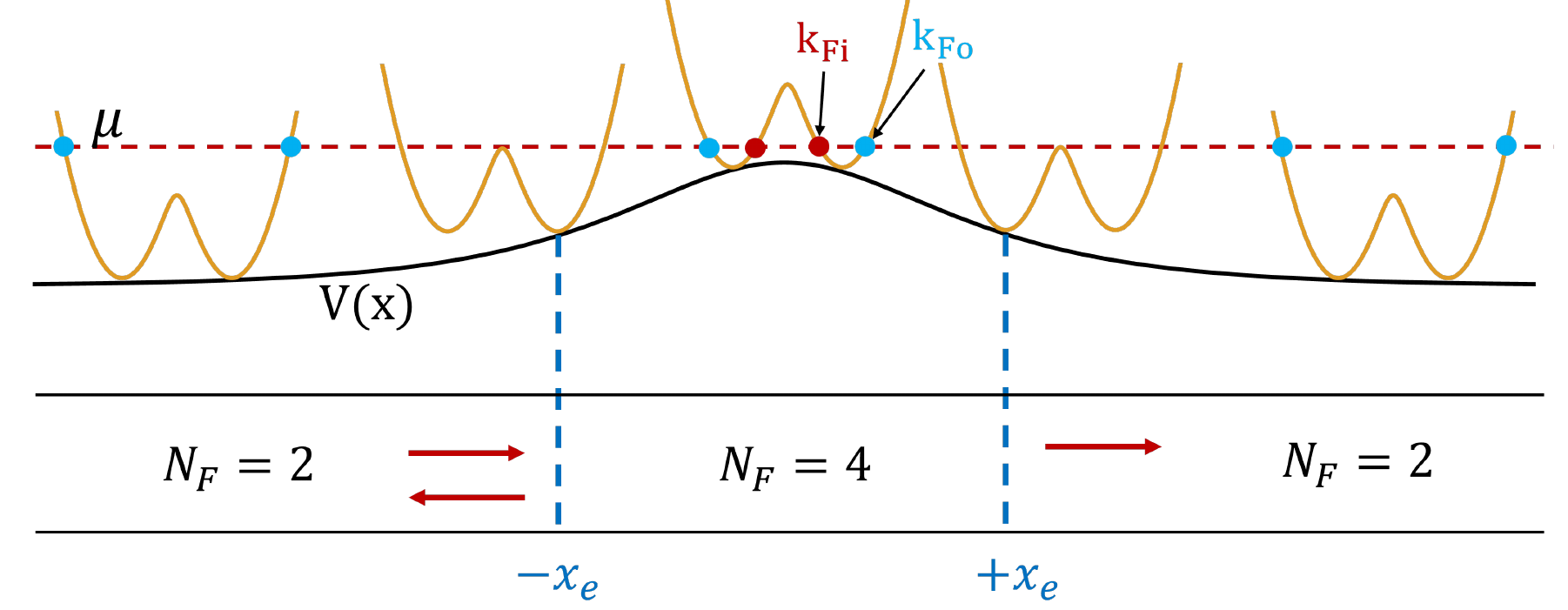}
\caption{\label{fig:Illustration_Fabry_Perot}
Schematic of the Fabry--P\'erot-like mechanism underlying magnetic-field-dependent conductance oscillations. 
As the potential $V(x)$ varies along $x$ (black curve), it vertically shifts the local dispersion, so that at a fixed energy (red dashed line at $\mu$) the number of Fermi points becomes position dependent. 
Far from the barrier center the spectrum has two Fermi points ($N_F=2$), whereas near the center a finite segment $|x|<x_e$ with four Fermi points ($N_F=4$) emerges. 
Scattering at the two interfaces $x=\pm x_e$ leads to partial reflections and hence Fabry--P\'erot-type interference for propagation through the $N_F=4$ region.}
\end{figure}

We compute the noninteracting magnetoconductance from the numerically obtained scattering matrix $S(\epsilon,B)$; the results are shown in Fig.~\ref{fig:NoninteractingG}.
Note that in the transition region the conductance exhibits pronounced oscillations as a function of magnetic field. 
To understand the origin of these oscillations, it is illuminating to consider the evolution of the single-particle spectrum along the quantum point contact, as depicted in Fig.~\ref{fig:Illustration_Fabry_Perot}.

As the electrostatic potential $V(x)$ varies along the transport direction, it shifts the local band dispersion vertically, so that the set of propagating states at a fixed energy (red dashed line at $\mu$) becomes position dependent.
The inset of Fig.~\ref{fig:NoninteractingG} complements this picture by showing how the dispersion evolves with magnetic field, which in turn controls whether  a region with four Fermi points   ($N_F=4$)   develops near the barrier top.
Far from the barrier center, $\mu$ intersects the local dispersion at two Fermi points ($N_F=2$), corresponding to the usual right- and left-moving modes. Near the barrier center, however, the hump in the dispersion allows $\mu$ to intersect the local bands at four Fermi points ($N_F=4$). This defines a central segment $|x|<x_e$ of length $2x_e$.

An electron incident from the outer region ($N_F=2$)  is partially reflected back at the interfaces $x=\pm x_e$, and partially  transmitted into the $N_F=4$ region.
Multiple reflections between the two interfaces thus make the segment $|x|<x_e$ an effective {\it Fabry--P\'erot} cavity~\cite{rainis2014conductance,kretinin2010multimode}.
The resulting interference produces conductance oscillations as the magnetic field is varied.
Constructive interference---and hence enhanced transmission---occurs when the phase accumulated across the cavity satisfies
\begin{equation}
\phi(B)\equiv \int_{-x_e(B)}^{x_e(B)} k_{Fi}(x;B)\,dx = n\pi,
\label{SI:FPphase}
\end{equation}
where $\pm k_{Fi}(x;B)$ are the Fermi wave vectors of the inner two Fermi points in the $N_F=4$ region and $n$ is an integer. Because both the cavity length $2x_e(B)$ and the local wave vector $k_{Fi}(x;B)$ depend on the magnetic field, the phase $\phi(B)$ varies with $B$, so the resonance condition is satisfied repeatedly, leading to conductance oscillations as a function of magnetic field. 
The agreement between the constructive-interference condition in Eq.~(\ref{SI:FPphase}) and the maxima of the conductance oscillations in Fig.~\ref{fig:NoninteractingG} is demonstrated in the Supplementary Material, Sec.~\ref{SIsec:G0}.

The  oscillations described above occur only in the vicinity of a conductance-step transition, i.e., when the relevant channel is partially transmitting. This is because the interference requires finite reflection amplitudes at the two interfaces $x=\pm x_e(B)$; deep inside a conductance plateau the transmission is nearly quantized and the reflection is strongly suppressed, so the oscillatory modulation becomes negligible.
At fixed chemical potential, sweeping $B$ drives a conductance step at $B=B_{\mathrm{trans}}(\mu)$. For the Fabry--P\'erot cavity to be present at the transition, the field must satisfy
\begin{equation}
B_c < B_{\mathrm{trans}}(\mu) < B_{\mathrm{hump}}.
\end{equation}
Here $B>B_c$ ensures that the asymptotic region has only two Fermi points.
The upper bound is set by $g\mu_B B_{\mathrm{hump}} = m_e^*\gamma^2$, the field at which the hump near $q=0$ disappears. Therefore, $B<B_{\mathrm{hump}}$ ensures that the hump is still present, thereby allowing a finite $N_F=4$ segment near the barrier top.
This criterion also clarifies the role of the chemical potential: if $\mu$ is too high, then $B_{\mathrm{trans}}(\mu)<B_c$; if it is too low, then $B_{\mathrm{trans}}(\mu)>B_{\mathrm{hump}}$. In either case, the condition for forming an $N_F=2\leftrightarrow4$ Fabry--P\'erot cavity is not satisfied.

\begin{figure}[t]
 \centering
\includegraphics[width=1\columnwidth]{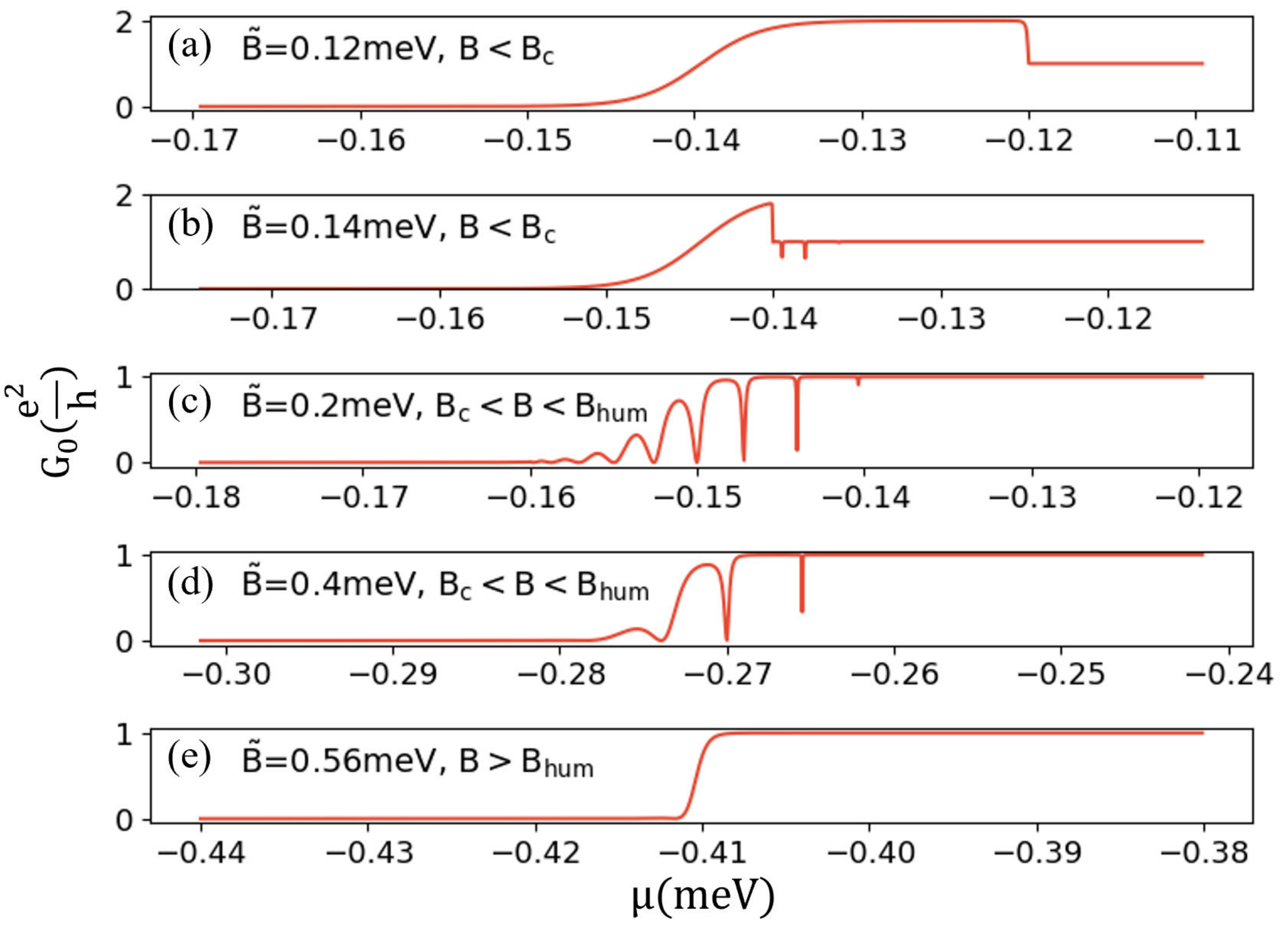}
\caption{
Noninteracting conductance $G_0$ as a function of chemical potential $\mu$ for $\boldsymbol{B}\perp\boldsymbol{\gamma}$ with $\hbar\gamma=4\,\mathrm{meV\cdot nm}$. Panels (a)–(e) correspond to different magnetic-field strengths: (a),(b) $B<B_c$; (c),(d) $B_c<B<B_{\mathrm{hump}}$; and (e) $B>B_{\mathrm{hump}}$. Fabry--P\'erot-like conductance oscillations occur in the intermediate regime (c),(d).
\label{fig:FBvsMu}}
\end{figure}

To further illustrate how the conductance oscillations depend on the chemical potential, Fig.~\ref{fig:FBvsMu} shows the noninteracting conductance $G_0(\mu)$ at several fixed magnetic fields for $\boldsymbol{B}\perp\boldsymbol{\gamma}$, covering several representative regimes.
 Complementary results showing the magnetic-field dependence of the conductance at different chemical potentials are provided in the Supplementary Material, Sec.~\ref{SIsec:G0}.
In the intermediate regime $B_c<B<B_{\mathrm{hump}}$, the conductance oscillates as a function of $\mu$ because varying $\mu$ changes both the cavity length $2x_e$ and the inner Fermi wave vector $k_{Fi}$, thereby tuning the Fabry--P\'erot phase. Outside this field window, no such cavity is formed, and the oscillations are absent.

\begin{figure}[t]
 \centering
\includegraphics[width=1\columnwidth]{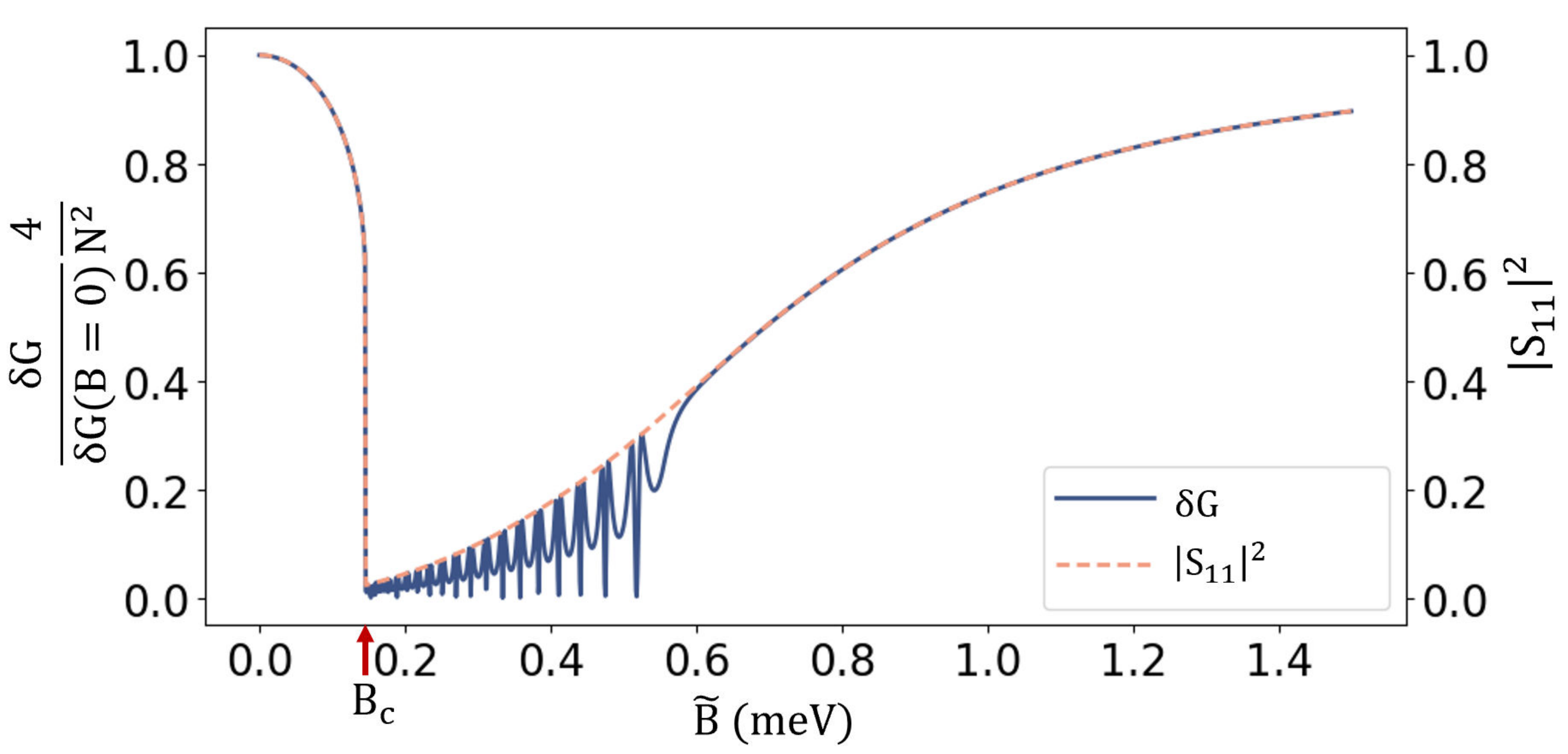}\caption{
Anomalous conductance $\delta G$ and spin overlap $|S_{11}|^2$ as a function of magnetic field.  
The setup and parameters are the same as in Fig.~\ref{fig:AnomalyVsChemicalPotential} and \ref{fig:NoninteractingG},
except that the chemical potential is always tuned to be at the barrier maximum $V\left(0\right)$ so that the channel of the lower subband is kept half open.  
The red arrow marks the critical field $B_c$. 
$N$ is the number of channels contributing to the transport.
\label{fig:FullB}
}
\end{figure}

\subsubsection{Interaction-induced conductance anomaly}
We next discuss the influence of the magnetic field on the anomalous conductance.
For $\boldsymbol{B}\perp\boldsymbol{\gamma}$ the interaction correction $\delta G$ is evaluated  analogously to the procedure used at zero field, with the scalar amplitudes $r_s,t_s$ replaced by the full matrices $r_{ss'}(\mu,B)$ and $t_{ss'}(\mu,B)$ that include inter-channel mixing. 
The resulting analytic expressions are given in the Supplementary Material, Sec.~\ref{SIsec:channelmixing}, and in practice we obtain $\delta G(B)$ by inserting the corresponding matrix elements extracted from the numerically obtained scattering matrix.

The anomalous conductance $\delta G$ exhibits a non-monotonic dependence on the field strength, as depicted in Fig.~\ref{fig:FullB}. 
In the weak-field regime $\delta G$ decreases with ${B}$, whereas at a stronger field it increases. 
To elucidate this behavior, we analyze the two limits separately and plot the full result in Fig.~\ref{fig:FullB}.

In the weak-field limit,
although 
scattering can occur between all Fermi points, the dominant contribution to the anomaly comes from scattering between $k_1^{+}$ and $k_1^{-}$, and between $k_2^{+}$ and $k_2^{-}$. 
The corresponding spin-overlap amplitude is defined as
$S_{ss'} \equiv \left\langle \chi_{s'}^{-}\middle|\chi_{s}^{+}\right\rangle$.
Therefore, the dominant contribution is governed by $|S_{11}|^2$ and $|S_{22}|^2$; in the present case, they are equal, $|S_{11}|^2=|S_{22}|^2$, and vary with the magnetic field.
 
As the magnetic field increases, the spin overlap decreases, leading to a suppression of the conductance anomaly.
When the magnetic field is increased to $B=B_c$, the two inner Fermi points merge and annihilate, reducing the number of Fermi points from $4$ to $2$, as shown in Fig.~\ref{fig:Dispersion}.
Further increase of the magnetic field enhances the spin overlap between the two remaining Fermi points and therefore strengthens the anomalous conductance.
As a result, the anomalous conductance $\delta G$ exhibits a non-monotonic dependence on  the magnetic field, as shown in Fig.~\ref{fig:FullB}.

The dashed curve in Fig.~\ref{fig:FullB} shows the spin overlap $|S_{11}|^2$, where the definitions of $k_1^{\pm}$ follow Fig.~\ref{fig:Dispersion}(a) for $B<B_c$ and Fig.~\ref{fig:Dispersion}(c) for $B>B_c$.  
This demonstrates that the non-monotonic behavior of $\delta G$ primarily originates from the non-monotonic dependence of the spin overlap on the magnetic field.
Another notable feature is that when the magnetic field strength increases across $B_c$, $\delta G$ exhibits oscillations. 
This is because the reflection oscillates with $B$, as shown in Fig.~\ref{fig:NoninteractingG}.

\section{Conclusion}
We have identified a universal interaction-driven mechanism that modifies the conductance of a partially open quantum channel. Even in clean systems with smooth barriers, backscattering  induces Friedel oscillations that generate singular corrections, maximized at half-open channels. 

When a magnetic field with a component perpendicular to the spin-orbit axis is applied, it creates a hump in the single-particle spectrum, 
leading to magnetic-field-dependent conductance oscillations originating from interference analogous to the  Fabry--P\'erot mechanism.
In addition, it affects the overlap of the spinors at different Fermi points,   resulting in a non-monotonic anomalous conductance.

Our analysis indicates that the conductance anomaly is a generic feature of adiabatic quantum point contacts, not limited to the first conductance step, and can also arise near the opening of higher channels.
In the multichannel regime, however, enhanced screening suppresses the effective interaction strength, making the anomaly weaker and harder to resolve experimentally.

\section*{acknowledgments}
We thank L.~Liu and P.~Ostrovsky for valuable discussions. D.~L. acknowledges support from the National Natural Science Foundation of China (Grant No.~12504003) and from the Guangdong Basic and Applied Basic Research Foundation (Grant No.~2026A1515010641).

\bibliography{manuscript}

\clearpage
\widetext
\section*{Supplementary Material for "Interaction-Induced Conductance Anomaly in a Partially Open Adiabatic Quantum Point Contact"}

\setcounter{section}{0}
\renewcommand{\thesection}{\Roman{section}}

\setcounter{figure}{0}
\setcounter{table}{0}
\setcounter{equation}{0}
\renewcommand{\thefigure}{S\arabic{figure}}
\renewcommand{\thetable}{S\arabic{table}}
\renewcommand{\theequation}{S\arabic{equation}}

\section{Eigenstates of P\"{o}schl-Teller Potential\label{SI:eigenstates}}

In the noninteracting case, the scattering states can be obtained by solving the Schr\"odinger equation:
\begin{equation}
\left(\tilde{H}-\tilde{\epsilon} \right)\Psi_a\left(x\right)=0,  \label{eq:tildeHequation}
\end{equation}
where $\tilde{\epsilon}$ denotes the eigenvalue and 
\begin{equation}
\tilde{H}=\frac{2m_e^*}{\hbar^2\alpha^2}h=\left[-\frac{d^2}{d\tilde{x}^2}-\frac{\lambda\left(\lambda-1\right)}{\cosh^{2}\left( \tilde{x}\right)}-\tilde{\mu}\right]\sigma_0-i\tilde{\boldsymbol{\gamma}}\cdot\boldsymbol{\sigma} \frac{d}{d\tilde{x}}+ g\tilde{\mu_B} \boldsymbol{B}\cdot\boldsymbol{\sigma}.\label{SI:fullHamiltonian}
\end{equation}
Here $h$ is the noninteracting Hamiltonian in Eq.~(\ref{eq:H0}) of the main text, and 
\begin{equation}
\tilde{x}=\alpha x,\ \tilde{\mu}=\frac{2m_e^{*}\mu}{\hbar^2\alpha^2},\ \tilde{\boldsymbol{\gamma}}=\frac{2m_e^{*}}{\hbar\alpha}{\boldsymbol{\gamma}},\ \tilde{\mu}_B=\frac{2m_e^{*}}{\hbar^2\alpha^2}{\mu_B}.
\end{equation}

We first address the case of $B=0$. In this case, the spins are polarized by $\tilde{\boldsymbol{\gamma}}\cdot\boldsymbol{\sigma}$. We can write the eigenstate as
\begin{equation}
\Psi_{1,\tilde{k}}\left(\tilde{x}\right)=e^{-i\frac{\tilde{\gamma}}{2} \tilde{x}}\psi_{1,\tilde{k}}\left(\tilde{x}\right)\left|\chi_{1}\right\rangle , \label{eq:smallPsi1}
\end{equation}
\begin{equation}
\Psi_{2,\tilde{k}}\left(\tilde{x}\right)=e^{i\frac{\tilde{\gamma}}{2} \tilde{x}}\psi_{2,\tilde{k}}\left(\tilde{x}\right)\left|\chi_{2}\right\rangle , \label{eq:smallPsi2}
\end{equation}
where $\left|\chi_{1}\right\rangle $ and $\left|\chi_{2}\right\rangle $ are two eigenvectors of $\ensuremath{\tilde{\boldsymbol{\gamma}}\cdot\boldsymbol{\sigma}}$ with eigenvalues $+\tilde{\gamma}$ and $-\tilde{\gamma}$ respectively. Substituting Eqs.~(\ref{eq:smallPsi1}) and (\ref{eq:smallPsi2}) into Eq.~(\ref{eq:tildeHequation}), one finds that $\psi_{1,\tilde{k}}\left(\tilde{x}\right)$ and $\psi_{2,\tilde{k}}\left(\tilde{x}\right)$ satisfy the same Schr\"odinger equation. Taking $\psi_{1,\tilde{k}}\left(\tilde{x}\right)=\psi_{2,\tilde{k}}\left(\tilde{x}\right)=\psi_{\tilde{k}}\left(\tilde{x}\right)$ and setting the eigenvalue to $\tilde{\epsilon}=\tilde{k}^2-\tilde{\mu}-\tilde{\gamma}^2/4$, with $\tilde{k}>0$, one obtains:
\begin{equation}
\left[-\frac{d^{2}}{d\tilde{x}^{2}}-\frac{\lambda\left(\lambda-1\right)}{\cosh^{2}\left(\tilde{x}\right)}-\tilde{k}^{2}\right]\psi_{\tilde{k}}\left(\tilde{x}\right)=0.  
\end{equation}
The solution of this equation is  
\begin{equation}
  \psi_{\tilde{k}}\left(\tilde{x}\right)=c_1 P^{i\tilde{k}}_{\lambda-1}\left(\tanh\left(\tilde{x}\right)\right)+c_2 Q^{i\tilde{k}}_{\lambda-1}\left(\tanh\left(\tilde{x}\right)\right),\label{eq:Generalsolution}
\end{equation}
where $P^{\nu}_{\mu}\left(z\right)$  and $Q^{\nu}_{\mu}\left(z\right)$ are associated Legendre functions of the first kind and the second kind, respectively. The two scattering states are:
\begin{equation}
  \psi_{\tilde{k}}^{L}\left(\tilde{x}\right)=c^{L} P^{i\tilde{k}}_{\lambda-1}\left(\tanh\left(\tilde{x}\right)\right),\label{LegendreL}
\end{equation}
where 
\begin{equation}
   c^L=\frac{\Gamma\left(1-i\tilde{k}-\lambda\right)\Gamma\left(-i\tilde{k}+\lambda\right)}{\Gamma\left(-i\tilde{k}\right)},
\end{equation}
and
\begin{equation}
   \psi_{\tilde{k}}^{R}\left(\tilde{x}\right)=c^{R}_1 P^{i\tilde{k}}_{\lambda-1}\left(\tanh\left(\tilde{x}\right)\right)+c^{R}_2 Q^{i\tilde{k}}_{\lambda-1}\left(\tanh\left(\tilde{x}\right)\right),\label{LegendreR}
\end{equation}
where 
\begin{equation}
c^{R}_1 =i{\tilde{k}}\frac{\Gamma\left(1-i\tilde{k}-\lambda\right)\Gamma\left(-i\tilde{k}+\lambda\right)\cosh\left(\pi\left(\tilde{k}-i\lambda\right)\right)}{\Gamma\left(1-i\tilde{k}\right)},
\end{equation}
\begin{equation}
c^{R}_2 =-\frac{{2}\Gamma\left(1+i\tilde{k}\right)\Gamma\left(-i\tilde{k}+\lambda\right)}{\Gamma\left(1-i\tilde{k}\right)\Gamma\left(i\tilde{k}\right)\Gamma\left(i\tilde{k}+\lambda\right)}.
\end{equation}
By taking the asymptotic forms of $\psi_{\tilde{k}}^{L,R}\left(\tilde{x}\right)$ at $\tilde{x}\rightarrow \pm \infty$, we can obtain the scattering states.
The asymptotic forms of $\psi_{{k}}^{L}\left({x}\right)$ are
\begin{equation}
  \psi_k^{L}\left(x\right)\sim e^{ikx}+\frac{\Gamma\left(i\tilde{k}\right)\Gamma\left(1-i\tilde{k}-\lambda\right)\Gamma\left(-i\tilde{k}+\lambda\right)}{\Gamma\left(\lambda\right)\Gamma\left(1-\lambda\right)\Gamma\left(-i\tilde{k}\right)}e^{-ikx}, \quad x \to -\infty, \label{eq:Lminus}
\end{equation}
\begin{equation}
\psi_k^{L}(x) \sim  \frac{\Gamma(1-i\tilde{k}-\lambda)\Gamma(-i\tilde{k}+\lambda)}{\Gamma(-i\tilde{k})\Gamma(1-i\tilde{k})}e^{ikx}, \quad x \to +\infty, \label{eq:Lplus}
\end{equation}
where $k=\alpha\tilde{k}>0$, $kx=\tilde{k}\tilde{x}$ and we have replaced the variable back to $x=\tilde{x}/\alpha$. The asymptotic forms of $\psi_{{k}}^{R}\left({x}\right)$ are
\begin{equation}
  \psi_k^{R}\left(x\right)\sim  e^{-ikx}-\frac{\Gamma\left(i\tilde{k}\right)\Gamma\left(1-i\tilde{k}-\lambda\right)\Gamma\left(-i\tilde{k}+\lambda\right)}{\Gamma\left(\lambda\right)\Gamma\left(1-\lambda\right)\Gamma\left(-i\tilde{k}\right)}e^{ikx}, \quad x \to +\infty, \label{eq:Rplus}
\end{equation}

\begin{equation}
\psi_k^{R}\left(x\right)\sim \frac{\Gamma(1-i\tilde{k}-\lambda)\Gamma(-i\tilde{k}+\lambda)}{\Gamma(-i\tilde{k})\Gamma(1-i\tilde{k})}e^{-ikx}, \quad x \to -\infty. \label{eq:Rminus}
\end{equation}
The reflection and transmission amplitudes can be extracted from the formulas above
\begin{equation}
\label{reflection}
r_{k}=\frac{\Gamma\left(i\tilde{k}\right)\Gamma\left(1-i\tilde{k}-\lambda\right)\Gamma\left(-i\tilde{k}+\lambda\right)}{\Gamma\left(\lambda\right)\Gamma\left(1-\lambda\right)\Gamma\left(-i\tilde{k}\right)}, 
\end{equation}
\begin{equation}
\label{transmission}
t_{k}=\frac{\Gamma\left(1-i\tilde{k}-\lambda\right)\Gamma\left(-i\tilde{k}+\lambda\right)}{\Gamma\left(-i\tilde{k}\right)\Gamma\left(1-i\tilde{k}\right)}.
\end{equation}
These are the reflection and transmission amplitudes used in the main text, Eqs.~(\ref{eq:rL}) and (\ref{eq:tL}). 

In fact, these results match the semiclassical (WKB) approximation.
For the simple case of no SOC and zero magnetic field, one easily finds the position of the turning points, where the kinetic and potential energies are equal.
The turning points $x_\pm (\epsilon)$ are determined by the condition $\cosh(\alpha x)= V(0)/\epsilon$ and have two solutions 
on the real axis $x_\pm=(\pm1/\alpha) \text{arccosh} \left(\sqrt{V(0)/\epsilon}\right)$  for $\epsilon<V(0)$ and on the imaginary axes
$x_\pm=(\pm i/\alpha) \text{arccosh} \left(\sqrt{V(0)/\epsilon}\right)$ for $\epsilon>V(0)$. 
The distance between the origin ($x=0$) and the turning points matches the scale $\ell(\epsilon)$.
In that regime, the reflection and transmission coefficients in Eqs.~(\ref{reflection}) and (\ref{transmission}) agree with the Campbell formula~\cite{BerryMount1972}. 

In summary, for the case $B=0$, the four  eigenstates have the following  asymptotic forms 
\begin{align}
\Psi_{1,k}^{L}\left(x\right)&=\psi_{1,k}^{L}\left(x\right)\left|\chi_{1}\right\rangle =e^{-i\frac{\gamma}{2} x}\psi_k^{L}\left(x\right)\left|\chi_{1}\right\rangle ,\\
\Psi_{2,k}^{L}\left(x\right)&=\psi_{2,k}^{L}\left(x\right)\left|\chi_{2}\right\rangle =e^{i\frac{\gamma}{2} x}\psi_k^{L}\left(x\right)\left|\chi_{2}\right\rangle , \\
\Psi_{1,k}^{R}\left(x\right)&=\psi_{1,k}^{R}\left(x\right)\left|\chi_{1}\right\rangle =e^{-i\frac{\gamma}{2} x}\psi_k^{R}\left(x\right)\left|\chi_{1}\right\rangle ,\\
\Psi_{2,k}^{R}\left(x\right)&=\psi_{2,k}^{R}\left(x\right)\left|\chi_{2}\right\rangle =e^{i\frac{\gamma}{2} x}\psi_k^{R}\left(x\right)\left|\chi_{2}\right\rangle ,
\end{align}
where $\psi_{s,k}^{L/R}$ and $\left|\chi_s\right\rangle$ are the spatial and spin parts of the wave function, respectively. 
Here we denote
\begin{align}
\psi_{k}^{L}\left(x\right)=\begin{cases}
e^{ik x}+r_{k}e^{-ikx}, & x<x_-,\\
t_{k}e^{ikx}, & x>x_+,
\end{cases}\label{eq:AssymL}
\end{align}
and
\begin{align}
\psi_{k}^{R}\left(x\right)=\begin{cases}
t_{k}e^{-ikx}, & x<x_-,\\
e^{-ik x}+r_{k}e^{ikx}, & x>x_+.
\end{cases}\label{eq:AssymR}
\end{align}
The corresponding scattering matrix is 
\begin{align}
S_{s,k}=
\left(\begin{array}{cc}
r^{L}_{s,k} & t^{R}_{s,k}\\
t^{L}_{s,k} & r^{R}_{s,k}
\end{array}\right)
=\left(\begin{array}{cc}
r_{k} & t_{k}\\
t_{k}  & r_{k} 
\end{array}\right),
\end{align}
with $S_{s,k}^{\dagger}S_{s,k}=I$.

\section{Friedel oscillations of the electron density}
Using asymptotic wave functions, the density can be expressed as 
\begin{align}
\rho\left(x\right)	
&=\sum_{s,k<k_{F}}\Psi_{s,k}^{R\dagger}\left(x\right)\Psi_{s,k}^{R}\left(x\right)+\Psi_{s,k}^{L\dagger}\left(x\right)\Psi_{s,k}^{L}\left(x\right) \label{SI:density}\\
&=2\sum_{k<k_{F}}\psi_{k}^{R\dagger}\left(x\right)\psi_{k}^{R}\left(x\right)+\psi_{k}^{L\dagger}\left(x\right)\psi_{k}^{L}\left(x\right). \label{SIeq:densityNoSOC}
\end{align}
The density can be obtained by substituting the asymptotics given in Eqs.~(\ref{eq:AssymL}) and (\ref{eq:AssymR})
\begin{equation}
\rho\left(x\right)= 
\begin{cases}
\frac{1}{2\pi}\int_{0}^{k_{F}}4\left[1+\text{Re}\left( r_{k}e^{-2ikx}\right)\right]dk=\frac{2k_F}{\pi}+\frac{2}{\pi}\int_{0}^{k_{F}}dk\text{Re}\left( r_{k}e^{-2ikx}\right) ,& x<0,\\
\frac{1}{2\pi}\int_{0}^{k_{F}}4\left[1+\text{Re}\left( r_{k}e^{2ikx}\right)\right]dk=\frac{2k_F}{\pi}+\frac{2}{\pi}\int_{0}^{k_{F}}dk\text{Re}\left( r_{k}e^{2ikx}\right) , & x>0.
\end{cases}
\end{equation}
At relatively large $\left|x\right|$, $e^{-2ikx}$ oscillates much faster than $r_k$ when $k$ changes. The above integral over $k$ gives rise to the correction to the density:  
\begin{equation}
\delta \rho \left(x\right)=
\begin{cases}
\frac{2}{\pi}\int_{0}^{k_{F}}\text{Re}\left( r_{k}e^{-2ikx}\right)dk \approx \frac{1}{2i\pi x}\left(r^{*}_{k_F}e^{2ik_F x}-r_{k_F}e^{-2ik_F x}\right) ,& x<0,\\
\frac{2}{\pi}\int_{0}^{k_{F}}\text{Re}\left( r_{k}e^{2ikx}\right)dk\approx \frac{1}{2i\pi x}\left(r_{k_F}e^{2ik_F x}-r^{*}_{k_F}e^{-2ik_F x}\right) , & x>0.
\end{cases} \label{SIeq:densitykF}
\end{equation}
This justifies the asymptotic form of the density oscillations in Eq.~(\ref{densityosc}) of the main text. The expression above is valid for $x$ such that  $V(x)\ll\mu$.

\section{Correction to the wave function}
The correction to the wave function due to the oscillating potential can be found with the Born approximation,
\begin{equation}
\delta\Psi_{m}\left(x\right)=\int dy\hat{G}_{m}\left(x,y\right)V_{H}\left(y\right)\Psi_{m}\left(y\right),\label{SI:LSequation}
\end{equation}
where $m$ stands for the labels $L/R,s,k$.  
The Green's function in Eq.~(\ref{SI:LSequation}) is defined as
\begin{equation}
\hat{G}_{m}\left(x,y\right)=\sum_{n}\frac{\Psi_{n}\left(x\right)\Psi_{n}^{\dagger}\left(y\right)}{E_{m}-E_{n}+i0^{+}},
\label{SI:Gfunction}
\end{equation}
where $0^{+}$ is an infinitesimal positive value.
Specifically, for the system in Section~\ref{SI:eigenstates}, namely Eq.~(\ref{SI:fullHamiltonian}) with $B=0$, the Green's function is:
\begin{align}
\hat{G}^{L}_{k,s}\left(x,y\right)=&\sum_{s',k'}\left[\frac{\Psi^{L}_{s',k'}\left(x\right)\Psi_{s',k'}^{L\dagger}\left(y\right)}{E_{L,s,k}-E_{L,s',k'}+i0^{+}}+\frac{\Psi^{R}_{s',k'}\left(x\right)\Psi_{s',k'}^{R\dagger}\left(y\right)}{E_{L,s,k}-E_{R,s',k'}+i0^{+}}\right]\\
=&\sum_{s',k'>0}\left[\frac{\psi^{L}_{s',k'}\left(x\right)\psi_{s',k'}^{L\dagger}\left(y\right)}{E_{L,s,k}-E_{L,s',k'}+i0^{+}}+\frac{\psi^{R}_{s',k'}\left(x\right)\psi_{s',k'}^{R\dagger}\left(y\right)}{E_{L,s,k}-E_{R,s',k'}+i0^{+}}\right]\hat{P}_{s'},
\end{align}
where $\psi^{L/R}_{s',k'}\left(x\right)$ is the spatial part of the wave function $\Psi^{L/R}_{s',k'}\left(x\right)$. $\hat{P}_{s'}=\left|\chi_{s'}\right\rangle\left\langle \chi_{s'}\right|$ is the projection operator. 
The scattering potential is not spin-dependent, $\hat{V}_{H}\left(y\right)=V_{H}\left(y\right)\sigma_0$. Eq.~(\ref{SI:LSequation}) can be rewritten as
\begin{equation}
\delta\Psi^{L}_{s,k}\left(x\right)=\int dy{G}^{L}_{s,k}\left(x,y\right){V}_{H}\left(y\right)\Psi^{L}_{s,k}\left(y\right),
\label{eq:deltaPhiSI}
\end{equation}
where
\begin{align}
G^{L}_{k,s}\left(x,y\right)=\sum_{k'>0}\left[\frac{\psi^{L}_{s,k'}\left(x\right)\psi_{s,k'}^{L*}\left(y\right)}{E_{L,s,k}-E_{L,s,k'}+i0^{+}}+\frac{\psi^{R}_{s,k'}\left(x\right)\psi_{s,k'}^{R*}\left(y\right)}{E_{L,s,k}-E_{R,s,k'}+i0^{+}}\right].
\end{align}
Using $E_{L,s,k}=E_{R,s,k}={\hbar^2 k_s^2}/{2m_e^{*}}-\mu-m_e^{*}{\gamma}^2/2$, where $s=\pm1$, $ k_1= k- m_e^*\gamma/\hbar$ and  $ k_2= k+ m_e^*\gamma/\hbar$, the Green's function can be written as 
\begin{align}
G^{L}_{k,s}\left(x,y\right)=-\frac{2m_e^{*}}{\hbar^2}\frac{1}{2\pi}\int_{0}^{\infty}\left[\frac{\psi^{L}_{s,k'}\left(x\right)\psi_{s,k'}^{L*}\left(y\right)}{\left(k'-k-i0^{+}\right)\left(k'+k+i0^{+}\right)}+\frac{\psi^{R}_{s,k'}\left(x\right)\psi_{s,k'}^{R*}\left(y\right)}{\left(k'-k-i0^{+}\right)\left(k'+k+i0^{+}\right)}\right]dk'.
\end{align}
Since both $k$ and $k'$ are positive,  $G^{L}_{k,s}\left(x,y\right)$ can be evaluated by taking the residue at $k'=k$. In the limit $x\rightarrow -\infty$,
\begin{align}
G^{L}_{s,k}\left(x,y\right)=
\frac{e^{ik_{s}^{-}x}}{i\hbar v_{k}}\begin{cases}
e^{-ik_{s}^{-}y}+r_{k}e^{-ik_{s}^{+}y}, & y<0,\\
t_{k} e^{-ik_{s}^{-}y}, & y>0,
\end{cases}\label{eq:GL}
\end{align}
where $v_{k}=\hbar k/m_e^{*}$.
$k_1^{\pm}=\pm k- k_{\gamma}$,  $ k_2^{\pm}=\pm k+ k_{\gamma}$ with $k_{\gamma}=m_e^{*} \gamma/\hbar$.

In the case of short range interacting potential 
$U(y-z)\simeq\beta\pi\hbar v_{F}\delta(y-z)$, with $\beta$ the dimensionless interaction strength, the Hartree potential ${V}_H\left(y\right)$ is
\begin{equation}
V_H\left(y\right)=\int_{-\infty}^{\infty}dz U\left(y-z\right)\delta\rho\left(z\right)=\beta\pi\hbar v_{F}\delta \rho\left(y\right).\label{eq:HartreePotential}
\end{equation} 
Substituting Eqs.~(\ref{eq:GL}) and (\ref{eq:HartreePotential}) into Eq.~(\ref{eq:deltaPhiSI}) yields the correction to the scattering state:
\begin{equation}
\delta\Psi_{s,k}^{L}\left(x\right)=\delta r_k e^{i k^{-}_{s} x}\left|\chi_{s}\right\rangle=\beta\mathcal{T}_k r_{k}\ln\left(\frac{1}{\left|k-k_F\right|L}\right)e^{i k^{-}_{s} x}\left|\chi_{s}\right\rangle,
\end{equation}
where $L$ is the ultraviolet cutoff length that is  set by the largest of the characteristic spatial scales:    
the effective width of the barrier  at the Fermi energy  $\ell(\epsilon_F)$, the Fermi wavelength far from the barrier 
$\lambda_F$,  and the characteristic scale of the short-range interaction. 

The resulting correction to the transmission is
\begin{align}
\delta \mathcal{T}_k=-\sum_{s}\left(\left|r_{s,k}+\delta r_{s,k}\right|^{2}-\left|r_{s,k}\right|^{2}\right)
\approx -4\text{Re}(r_{k}^{*}\delta r_{k})
=-4\beta\mathcal{T}_k \left(1-\mathcal{T}_k\right)\ln\left(\frac{1}{\left|k-k_F\right|L}\right),
\end{align}
which corresponds to  Eq.~(\ref{eq:delta_T}) of the main text. 
\begin{figure}[!htbp]
\includegraphics[width=1\columnwidth]{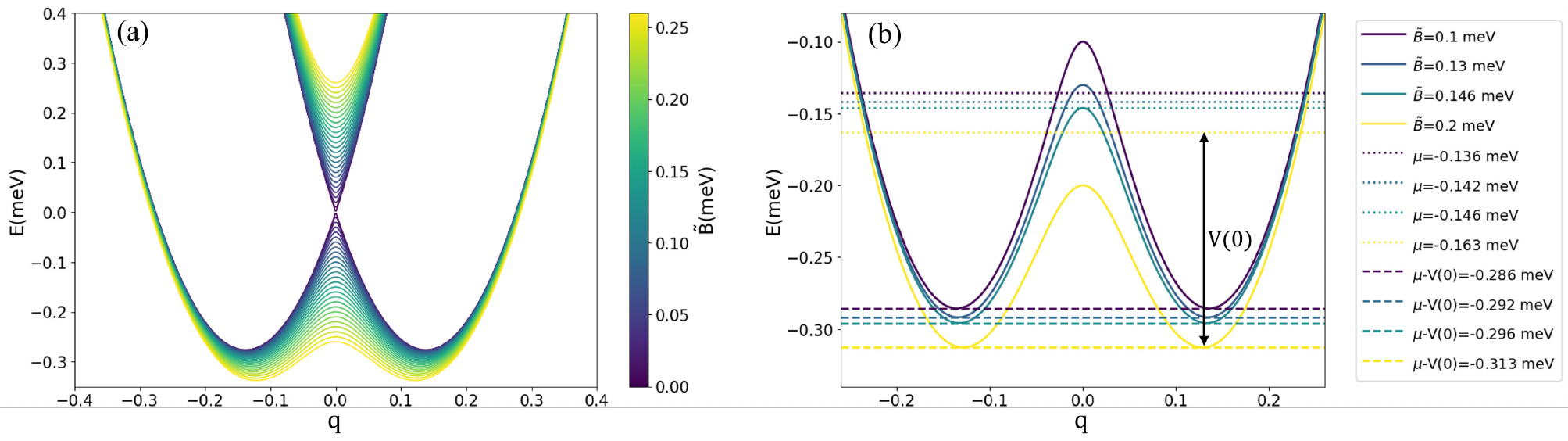}\caption{\label{fig:SIdispersionvsB}
(a) Energy spectrum far from the barrier for different magnetic fields $B$, with the field oriented perpendicular to the SOC axis.
(b) Energy spectra at four representative magnetic fields.
In each case, the band bottom aligns with  $\mu-V(0)$ (dashed lines), corresponding to a half-open channel.
The four chemical potentials (dotted lines) are chosen to match those used in Fig.~\ref{fig:SIG0vsB}, where the conductance evolves from $0$ to a quantized value as the magnetic field increases.
Each spectrum therefore corresponds to the magnetic field at which the channel becomes half open for the respective chemical potential.
}
\end{figure}

\section{Interplay of Perpendicular Magnetic Field and Spin-Orbit Coupling}
Next,  we consider the case where the magnetic field $\boldsymbol{B}$  is perpendicular to the direction of the SOC field $\boldsymbol{\gamma}$. 
The spectrum far away from the barrier is 
\begin{equation}
\label{spectrum}
E_{\pm}(q)=\frac{\hbar^2 q^2}{2m_e^*}\pm\sqrt{\hbar^2\gamma^2q^2+g^2{\mu_B}^2B^2},
\end{equation}
which is $B$-dependent, as shown in Fig.~\ref{fig:SIdispersionvsB}(a). 
Note that  the energy of the lower branch  $E_-(q)$  for small magnetic fields $g\mu_B B < \gamma^2 m_e^*$  is a non-monotonic function of $q$.
It has  a maximum at $q=0$  and two minima at  $\hbar q_{\rm min}=\pm \sqrt{(m_e^*\gamma)^2-(g\mu_B B/\gamma)^2}$, 
with the value at the minima 
$E(q_{\rm min})=-(g\mu_B B)^2/(2m_e^*\gamma^2)-m_e^*\gamma^2/2$.

The magnetic field has two main effects. First, together with SOC, it modifies the spectrum, producing a non-monotonic energy–momentum relation with a local maximum (hump) at the origin.
This leads to oscillations in the noninteracting conductance as the channel opens, through Fabry--P\'erot–type interference.
Second, it alters the spinor polarization, changing the overlap between spinors at different Fermi points. This governs the Friedel-oscillation–induced anomaly, resulting in a non-monotonic dependence of the anomalous conductance on the magnetic field.

\begin{figure}[!htbp]
\includegraphics[width=0.9\columnwidth]{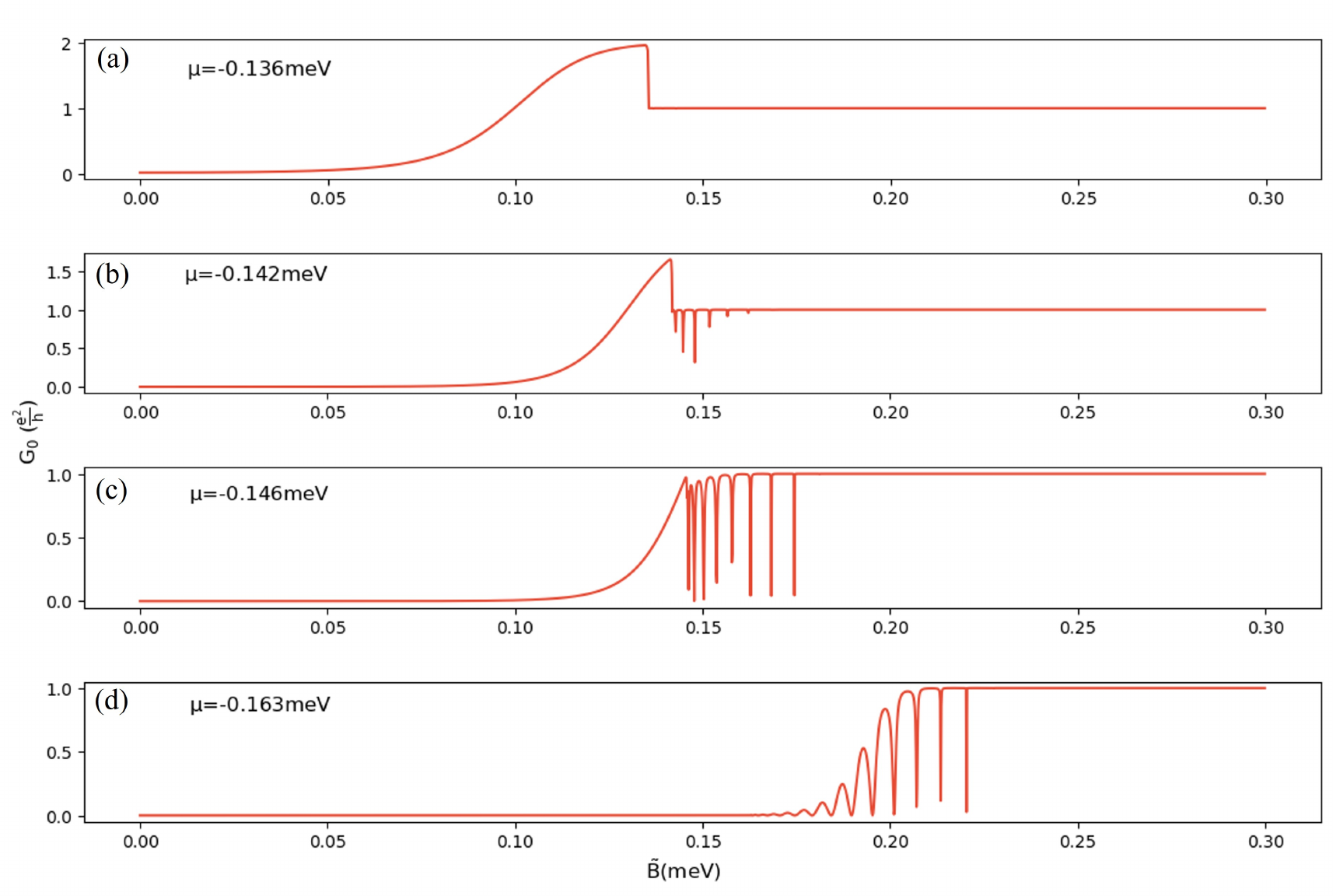}\caption{\label{fig:SIG0vsB}
Noninteracting conductance $G_0$ as a function of magnetic field strength $\tilde{B}=g\mu_BB$, under four different values of the chemical potential. As the absolute value of the chemical potential increases, the magnetic field at which the conductance step occurs also increases, corresponding to a change in the number of channels. When  $B \gtrsim B_c$ ($g\mu_B B_c=0.146$meV), the conductance changes from $e^2/h$ to $2e^2/h$, and conductance oscillations occur within the transition region.
}
\end{figure}
We first consider the noninteracting level.
\subsection{Noninteracting conductance \label{SIsec:G0}}
In the case of the perpendicular magnetic field and SOC, the noninteracting conductance $G_0$ as a function of the magnetic field with different chemical potential is shown in Fig.~\ref{fig:SIG0vsB}.
The data reveal distinct oscillations in  $G_0$ within the regime where the conductance evolves toward the $e^2/h$ plateau (where a single conducting channel opens).
Note that by increasing the absolute value of the chemical potential $\mu$, transitions to $e^2/h$ happen at larger $B$ and therefore the oscillations happen at larger $B$. The oscillations occur only for values of $B$ with $B \gtrsim B_c$ ($g\mu_B B_c = m_e^*\gamma^2-\sqrt{2m_e^*V(0)}\,\gamma$). $B_c$ is the critical field at which the number of Fermi points far from the barrier decreases from four to two, as shown in Fig.~\ref{fig:Dispersion} of the main text. 
In Fig.~\ref{fig:SIdispersionvsB}(b), we mark the positions of the four chemical potentials with dotted lines. We also indicate the position of $\mu - V(0)$ with dashed lines, which represents the kinetic energy at the top of the potential barrier. A conduction channel opens when the band bottom crosses $\mu - V(0)$. 
Therefore, in Fig.~\ref{fig:SIdispersionvsB}(b) we plot the corresponding band diagrams for the opening of the four channels and mark the magnetic field values associated with each of the four spectra.

\begin{figure}[h]
\includegraphics[width=0.9\columnwidth]{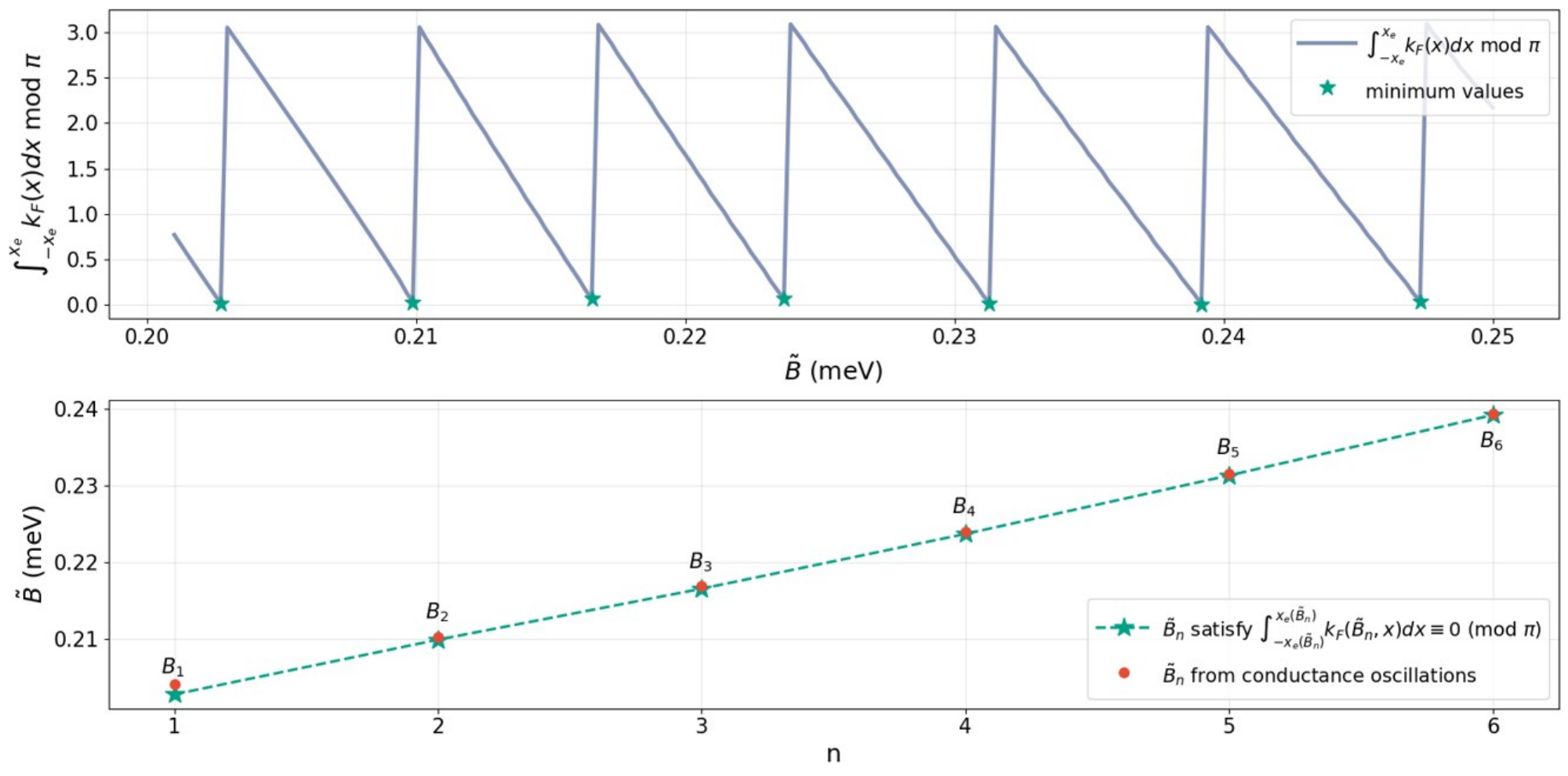}\caption{\label{fig:SIFittingBn}
(a) Phase $\phi(B)$ from Eq.~(\ref{SI:FPphase}) modulo $\pi$ as a function of $\tilde{B}$. The minima points, marked by green stars, indicate the fields $\tilde{B}_n$ satisfying the maximum transmission condition in Eq.~(\ref{SI:FPphase}) .
(b) Magnetic fields $\tilde{B}_n$ corresponding to the conductance maxima, extracted from the conductance oscillation pattern in Fig.~\ref{fig:SIG0vsB}(d), plotted versus the integer $n$ (denoting the $n$-th minimum), shown as red dots. Green stars represent $\tilde{B}_n$ from the minima in (a).
}
\end{figure}

The conductance oscillations in Fig.~\ref{fig:SIG0vsB} can be understood from the Fabry--P\'erot-like interference mechanism discussed in the main text and illustrated in Fig.~\ref{fig:Illustration_Fabry_Perot}. Let us consider the case shown in Fig.~\ref{fig:SIG0vsB}(d). As the magnetic field approaches $0.2$ meV, the energy of the local band minimum near the top of the potential barrier approaches $\mu - V(0)$, allowing electrons to tunnel through and open the conduction channel.

In this regime, the chemical potential $\mu$ intersects the local dispersion at two Fermi points ($N_F=2$) far from the barrier and at four Fermi points ($N_F=4$) near the barrier center, thereby creating a central region of length $2x_e$. Electrons transmitted into this region can populate the inner modes with Fermi wave vectors $\pm k_{Fi}$ and undergo multiple reflections between the two interfaces at $\pm x_e$, forming an effective Fabry--P\'erot cavity. Constructive interference -- and hence maximum transmission -- occurs when the accumulated phase satisfies
\begin{equation}
\phi\left(B\right)=\int_{-x_e}^{x_e} k_{Fi}\left(x\right)dx=n\pi,
\label{SI:FPphase}
\end{equation}
where $\pm k_{Fi}$ denote the inner two Fermi wave vectors within the $N_F=4$ region.
As the magnetic field increases, both $x_e$ and $k_{Fi}$ vary, so the resonance condition is satisfied repeatedly. This gives rise to the oscillatory conductance as a function of magnetic field.

In Fig.~\ref{fig:SIFittingBn}(a), we plot the phase $\phi(B)=\int_{-x_e}^{x_e} k_{Fi}\left(x\right)dx$ modulo $\pi$ as a function of $B$.
The minima of $\phi(B)$ determine the values $\tilde{B}_n$ at which the condition for maximum transmission in Eq.~(\ref{SI:FPphase}) is satisfied.
In Fig.~\ref{fig:SIFittingBn}(b), we compare these $\tilde{B}_n$ with the magnetic fields at which the conductance reaches its maximum, extracted from Fig.~\ref{fig:SIG0vsB}(d).
As shown, the two sets of values agree very well.
In Fig.~\ref{fig:SIG0vsB}(a), since the channel opens at $B < B_c$, the number of Fermi points far from the central region is four, no Fabry--P\'erot-type cavity is formed, and therefore no conductance oscillations occur.

\subsection{Anomalous conductance with scattering-induced inter-channel mixing\label{SIsec:channelmixing}}
\newcommand{\Dk}[2]{\Delta k_{#1#2}}
\newcommand{\Sov}[2]{S_{#1#2}}
\newcommand{\rL}[2]{r_{#1#2}^{L}}
\newcommand{\rR}[2]{r_{#1#2}^{R}}

\begin{figure}[h]
\includegraphics[width=0.7\columnwidth]{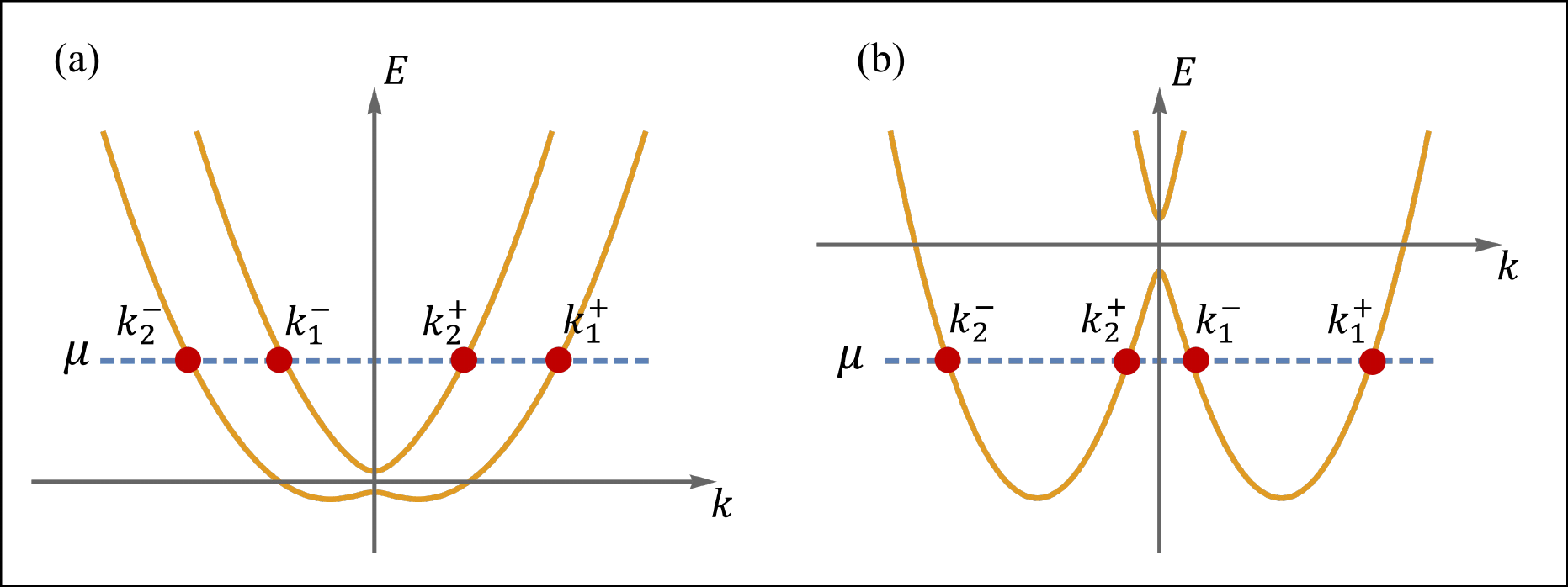}\caption{\label{fig:SIdispersion}
Electron dispersion away from the barrier with weak SOC (a) and strong SOC (b). A weak magnetic field is applied, opening a helical gap at $k=0$.  
}
\end{figure}

Eq.~(\ref{eq:delta_T}) in the main text corresponds to the result in the absence of mixing between the two spin-polarized channels. When the magnetic field possesses a component perpendicular to the spin-orbit coupling axis, scattering induces inter-channel mixing. Below, we derive the expression for the anomalous conductance in this case.
Fig.~\ref{fig:SIdispersion}(a) and (b) show the spectrum away from the barrier with weak SOC and strong SOC, respectively. At zero magnetic field, scattering only occurs between $k_1^+$ and $k_1^-$, and similarly between $k_2^+$ and $k_2^-$. Backward scattering between $k_1^+$ and $k_2^-$, as well as between $k_2^+$ and $k_1^-$, is forbidden by time-reversal symmetry. Increasing the magnetic field induces scattering between $k_1^+$ and $k_2^-$, and likewise between $k_2^+$ and $k_1^-$. The asymptotic form of the scattering states can be written as
\begin{align}
\Psi_{1,k}^{L}\left(x\right)&=
\begin{cases}
e^{ik_{1}^{+}x}\left|\chi_{1}^{+}\right\rangle +r_{11}^{L}e^{ik_{1}^{-}x}\left|\chi_{1}^{-}\right\rangle +r_{21}^{L}e^{ik_{2}^{-}x}\left|\chi_{2}^{-}\right\rangle , & x \ll -\ell(\epsilon),\\
t_{11}^{L}e^{ik_{1}^{+}x}\left|\chi_{1}^{+}\right\rangle +t_{21}^{L}e^{ik_{2}^{+}x}\left|\chi_{2}^{+}\right\rangle,  & x \gg \ell(\epsilon),
\end{cases}\label{SI:psi1L}
\\
\Psi_{2,k}^{L}\left(x\right)&=
\begin{cases}
e^{ik_{2}^{+}x}\left|\chi_{2}^{+}\right\rangle +r_{12}^{L}e^{ik_{1}^{-}x}\left|\chi_{1}^{-}\right\rangle +r_{22}^{L}e^{ik_{2}^{-}x}\left|\chi_{2}^{-}\right\rangle,  & x \ll -\ell(\epsilon),\\
t_{12}^{L}e^{ik_{1}^{+}x}\left|\chi_{1}^{+}\right\rangle +t_{22}^{L}e^{ik_{2}^{+}x}\left|\chi_{2}^{+}\right\rangle ,  & x \gg \ell(\epsilon),
\end{cases}
\\
\Psi_{1,k}^{R}\left(x\right)&=
\begin{cases}
t_{11}^{R}e^{ik_{1}^{-}x}\left|\chi_{1}^{-}\right\rangle +t_{21}^{R}e^{ik_{2}^{-}x}\left|\chi_{2}^{-}\right\rangle ,  & x \ll -\ell(\epsilon),\\
e^{ik_{1}^{-}x}\left|\chi_{1}^{-}\right\rangle +r_{11}^{R}e^{ik_{1}^{+}x}\left|\chi_{1}^{+}\right\rangle +r_{21}^{R}e^{ik_{2}^{+}x}\left|\chi_{2}^{+}\right\rangle ,  & x \gg \ell(\epsilon),
\end{cases}
\\
\Psi_{2,k}^{R}\left(x\right)&=
\begin{cases}
t_{12}^{R}e^{ik_{1}^{-}x}\left|\chi_{1}^{-}\right\rangle +t_{22}^{R}e^{ik_{2}^{-}x}\left|\chi_{2}^{-}\right\rangle  ,  & x \ll -\ell(\epsilon),\\
e^{ik_{2}^{-}x}\left|\chi_{2}^{-}\right\rangle +r_{22}^{R}e^{ik_{2}^{+}x}\left|\chi_{2}^{+}\right\rangle +r_{12}^{R}e^{ik_{1}^{+}x}\left|\chi_{1}^{+}\right\rangle  ,  & x \gg \ell(\epsilon).
\end{cases}\label{SI:psi2R}
\end{align}
Here, $k_1^{+},k_2^{+},k_1^{-},k_2^{-}$ represent the $k$-values on each energy band when the energy takes the same value $\epsilon$, as shown in Fig.~\ref{fig:SIdispersion}, and $k_{1,F}^{+},k_{2,F}^{+},k_{1,F}^{-},k_{2,F}^{-}$ denote the corresponding $k$-values when $\epsilon$ equals the Fermi energy. We represent the spin state at $k_{s}^{\pm}$ in Dirac notation $\left|\chi_{s}^{\pm}\right\rangle$.
The corresponding scattering matrix is written as
\begin{equation}
 S=\left(\begin{array}{cc}
r_{L} & t_{R}\\
t_{L} & r_{R}
\end{array}\right),
\text{ with }
r_{L/R}=\left(\begin{array}{cc}
r_{11}^{L/R} & r_{12}^{L/R}\\
r_{21}^{L/R} & r_{22}^{L/R}
\end{array}\right)
\text{ and }
t_{L/R}=\left(\begin{array}{cc}
t_{11}^{L/R} & t_{12}^{L/R}\\
t_{21}^{L/R} & t_{22}^{L/R}
\end{array}\right).
\label{SI:Smatrix}
\end{equation}

The corresponding density constructed from the wave functions in Eq.~(\ref{SI:psi1L})--(\ref{SI:psi2R}) is
\begin{align}
\rho_{k}\left(x\right)
&=
\sum_{s=1,2}\Psi_{s,k}^{L\dagger}\left(x\right)\Psi_{s,k}^{L}\left(x\right)
+\Psi_{s,k}^{R\dagger}\left(x\right)\Psi_{s,k}^{R}\left(x\right)\\
&=
\begin{cases}
4+2\mathrm{Re}\!\left[
\ r_{11}^{L*}e^{i\Delta k_{11}x}S_{11}
+r_{22}^{L*}e^{i\Delta k_{22}x}S_{22}
+r_{21}^{L*}e^{i\Delta k_{12}x}S_{12}
+r_{12}^{L*}e^{i\Delta k_{21}x}S_{21}
\right], & x \ll -\ell(\epsilon),\\[1mm]
4+2\mathrm{Re}\!\left[
\ r_{11}^{R}e^{i\Delta k_{11}x}S_{11}
+r_{22}^{R}e^{i\Delta k_{22}x}S_{22}
+r_{12}^{R}e^{i\Delta k_{12}x}S_{12}
+r_{21}^{R}e^{i\Delta k_{21}x}S_{21}
\right], & x \gg \ell(\epsilon).
\end{cases}
\end{align}
where the unitary property of the scattering matrix in Eq.~(\ref{SI:Smatrix}) has been used:  $S^{\dagger}S=I$, and we have defined, for $s,s'=1,2$,
\begin{equation}
\Delta k_{ss'} \equiv k_{s}^{+}-k_{s'}^{-},\qquad
S_{ss'} \equiv \left\langle \chi_{s'}^{-}\middle|\chi_{s}^{+}\right\rangle .
\end{equation}
Following Eq.~(\ref{SI:density}), we can obtain the density $\rho\left(x\right)=\bar{\rho}\left(x\right)+\delta\rho\left(x\right)$, with
\begin{align}
\delta\rho\left(x\right)
\approx
\begin{cases}
\frac{1}{2\pi x}\mathrm{Im}\!\left[
r_{11}^{L*}e^{i\Delta k^{F}_{11}x}S_{11}
+r_{22}^{L*}e^{i\Delta k^{F}_{22}x}S_{22}
+r_{21}^{L*}e^{i\Delta k^{F}_{12}x}S_{12}
+r_{12}^{L*}e^{i\Delta k^{F}_{21}x}S_{21}
\right],  & x\ll -\ell(\epsilon),\\[1mm]
\frac{1}{2\pi x}\mathrm{Im}\!\left[
r_{11}^{R}e^{i\Delta k^{F}_{11}x}S_{11}
+r_{22}^{R}e^{i\Delta k^{F}_{22}x}S_{22}
+r_{12}^{R}e^{i\Delta k^{F}_{12}x}S_{12}
+r_{21}^{R}e^{i\Delta k^{F}_{21}x}S_{21}
\right],  & x\gg\ell(\epsilon),
\end{cases}
\end{align}
where we denote the corresponding Fermi wave vectors as $k_{1,F}^{+},k_{2,F}^{+},k_{1,F}^{-},k_{2,F}^{-}$ (i.e., $k_{s,F}^{\pm}\equiv k_{s}^{\pm}(\epsilon_F)$), and correspondingly $\Delta k^{F}_{ss'}\equiv k_{s,F}^{+}-k_{s',F}^{-}$. The above expression is valid for $x$ such that  $V(x)\ll \mu$.

The correction to the wave function due to the oscillating potential can be obtained using Eq.~(\ref{SI:LSequation}), where the Green's function is obtained by substituting the wave functions in Eqs.~(\ref{SI:psi1L})--(\ref{SI:psi2R}) into Eq.~(\ref{SI:Gfunction}).
The corrections to the wave functions $\Psi_{1,k}^{L}\left(x\right)$ and $\Psi_{2,k}^{L}\left(x\right)$ are
\begin{align}
\delta\Psi_{1,k}^{L}\left(x\right)=&\delta r^L_{11}e^{ik_{1}^{-}x}\left|\chi_{1}^{-}\right\rangle+\delta r^L_{21} e^{ik_{2}^{-}x}\left|\chi_{2}^{-}\right\rangle, \\
\delta\Psi_{2,k}^{L}\left(x\right)=&\delta r^L_{12}e^{ik_{1}^{-}x}\left|\chi_{1}^{-}\right\rangle+\delta r^L_{22} e^{ik_{2}^{-}x}\left|\chi_{2}^{-}\right\rangle,
\end{align}
where the corrections to the reflection matrix elements are
\begin{align}
\delta r^L_{11}=&\frac{V_0}{4\pi \hbar v_{k}}
\left\{ \ln\left(\frac{1}{\left|\Delta k_{11}-\Delta k^{F}_{11}\right|L}\right)\left|S_{11}\right|^{2}\right.  \nonumber \\  
&
\left[r_{11}^{L}\left(1-r_{11}^{L}r_{11}^{L*}\right)-r_{11}^{R*}t_{11}^{L}t_{11}^{R}+r_{22}^{L}-r_{22}^{L*}r_{11}^{L}r_{11}^{L}
-r_{22}^{R*}t_{11}^{L}t_{11}^{R}
-r_{11}^{L*}r_{21}^{L}r_{12}^{L}-r_{22}^{L*}r_{12}^{L}r_{21}^{L}
-r_{11}^{R*}t_{21}^{L}t_{12}^{R}-r_{22}^{R*}t_{21}^{L}t_{12}^{R}
\right]  \nonumber \\
&+\ln\left(\frac{1}{\left|\Delta k_{12}-\Delta k^{F}_{12}\right|L}\right)\left|S_{12}\right|^{2}\left(-r_{21}^{L*}r_{21}^{L}r_{11}^{L}-r_{12}^{R*}t_{11}^{L}t_{12}^{R}\right)\nonumber\\
&\left.+\ln\left(\frac{1}{\left|\Delta k_{21}-\Delta k^{F}_{21}\right|L}\right)\left|S_{21}\right|^{2}\left(-r_{12}^{L*}r_{11}^{L}r_{12}^{L}-r_{21}^{R*}t_{21}^{L}t_{11}^{R}\right)
\right\},\label{SI:dr11}
\end{align}

\begin{align}
\delta r^L_{21}=&\frac{V_0}{4\pi \hbar v_{k}}
\left\{
\ln\left(\frac{1}{\left|\Delta k_{11}-\Delta k^{F}_{11}\right|L}\right)\left|S_{11}\right|^{2}\right. \nonumber  \\ 
&
\left(-r_{11}^{L}r_{21}^{L}r_{11}^{L*}-r_{11}^{L*}r_{21}^{L}r_{22}^{L}-r_{22}^{L*}r_{11}^{L}r_{21}^{L}-r_{22}^{L*}r_{21}^{L}r_{22}^{L}
 -r_{11}^{R*}t_{11}^{L}t_{21}^{R}-r_{11}^{R*}t_{21}^{L}t_{22}^{R}-r_{22}^{R*}t_{11}^{L}t_{21}^{R}-r_{22}^{R*}t_{21}^{L}t_{22}^{R}\right) \nonumber \\
&+\ln\left(\frac{1}{\left|\Delta k_{12}-\Delta k^{F}_{12}\right|L}\right)
\left|S_{12}\right|^{2}\left(r_{21}^{L}-r_{21}^{L}r_{21}^{L*}r_{21}^{L}-r_{12}^{R*}t_{11}^{L}t_{22}^{R}\right) \nonumber\\
&\left.
+\ln\left(\frac{1}{\left|\Delta k_{21}-\Delta k^{F}_{21}\right|L}\right)
\left|S_{21}\right|^{2}\left(-r_{12}^{L*}r_{11}^{L}r_{22}^{L}-r_{21}^{R*}t_{21}^{L}t_{21}^{R}\right)\right\},
\end{align}

\begin{align}
\delta r^L_{12}=&\frac{V_0}{4\pi \hbar v_{k}}
\left\{
\ln\left(\frac{1}{\left|\Delta k_{11}-\Delta k^{F}_{11}\right|L}\right)\left|S_{11}\right|^{2} \right. \nonumber \\ 
&
\left(-r_{22}^{L*}r_{12}^{L}r_{22}^{L}-r_{22}^{L*}r_{12}^{L}r_{11}^{L}-r_{11}^{L*}r_{22}^{L}r_{12}^{L}-r_{11}^{L*}r_{12}^{L}r_{11}^{L}
-r_{22}^{R*}t_{22}^{L}t_{12}^{R}-r_{22}^{R*}t_{12}^{L}t_{11}^{R}-r_{11}^{R*}t_{12}^{L}t_{11}^{R}-r_{11}^{R*}t_{22}^{L}t_{12}^{R}\right)\nonumber\\
&+\ln\left(\frac{1}{\left|\Delta k_{21}-\Delta k^{F}_{21}\right|L}\right)
\left|S_{21}\right|^{2}\left(r_{12}^{L}-r_{12}^{L}r_{12}^{L*}r_{12}^{L}-r_{21}^{R*}t_{22}^{L}t_{11}^{R}\right)\nonumber \\
&\left.+\ln\left(\frac{1}{\left|\Delta k_{12}-\Delta k^{F}_{12}\right|L}\right)
\left|S_{12}\right|^{2}\left(-r_{21}^{L*}r_{22}^{L}r_{11}^{L}-r_{12}^{R*}t_{12}^{L}t_{12}^{R}\right)\right\},
\end{align}

\begin{align}
\delta r^L_{22}=&\frac{V_0}{4\pi \hbar v_{k}}
\left\{
\ln\left(\frac{1}{\left|\Delta k_{22}-\Delta k^{F}_{22}\right|L}\right)\left|S_{22}\right|^{2}  
\right.\nonumber\\ 
&
\left[r_{22}^{L}\left(1-r_{22}^{L}r_{22}^{L*}\right)-r_{22}^{R*}t_{22}^{L}t_{22}^{R}
+r_{11}^{L}-r_{11}^{L*}r_{22}^{L}r_{22}^{L}-r_{11}^{R*}t_{22}^{L}t_{22}^{R}
-r_{22}^{L*}r_{12}^{L}r_{21}^{L}-r_{11}^{L*}r_{12}^{L}r_{21}^{L}
-r_{22}^{R*}t_{12}^{L}t_{21}^{R}-r_{11}^{R*}t_{12}^{L}t_{21}^{R}
\right]\nonumber\\
&+\ln\left(\frac{1}{\left|\Delta k_{12}-\Delta k^{F}_{12}\right|L}\right)\left|S_{12}\right|^{2}\left(-r_{21}^{L*}r_{22}^{L}r_{21}^{L}-r_{12}^{R*}t_{12}^{L}t_{22}^{R}\right)\nonumber\\
&\left.
+\ln\left(\frac{1}{\left|\Delta k_{21}-\Delta k^{F}_{21}\right|L}\right)\left|S_{21}\right|^{2}\left(-r_{12}^{L*}r_{12}^{L}r_{22}^{L}-r_{21}^{R*}t_{22}^{L}t_{21}^{R}\right)
\right\}.\label{SI:dr22}
\end{align}
In deriving Eqs.~(\ref{SI:dr11})--(\ref{SI:dr22}), we approximate the density of states at $k_{1}^{+}$ and $k_{2}^{+}$
to be equal, or equivalently, $v_{k_{2}^{+}}\approx v_{k_{1}^{+}}=v_{k}$, 
where $v_{k_{1}^{+}}$ and $v_{k_{2}^{+}}$ are the group velocities at $k_{1}^{+}$ and $k_{2}^{+}$, respectively. We also used the property $\left|S_{11}\right|^{2}=\left|S_{22}\right|^{2}$.

The transmission coefficient is given by 
\begin{align}
\mathcal{T}_k=2-\sum_{a=1,2}\sum_{b=1,2}\left|r^L_{ab}\right|^{2} ,
\end{align}
and its correction reads
\begin{align}
\delta \mathcal{T}_k=-\sum_{a=1,2}\sum_{b=1,2}\left(\left|r^L_{ab}+\delta r^L_{ab}\right|^{2}-\left|r^L_{ab}\right|^{2}\right)
\approx -2\sum_{a=1,2}\sum_{b=1,2}\text{Re}(r_{ab}^{L*}\delta r^L_{ab}). \label{SIeq:deltT}
\end{align}
At finite temperature, the anomalous conductance is obtained as
\begin{align}
\delta G = \frac{e^2}{h} \int \delta T(\epsilon) \left( -\frac{\partial f_{\epsilon}}{\partial \epsilon} \right) d\epsilon,\label{SIeq:GT}
\end{align}
where $f_{\epsilon}$ is the Fermi-Dirac distribution.

\begin{figure}[!t]
\includegraphics[width=0.8\columnwidth]{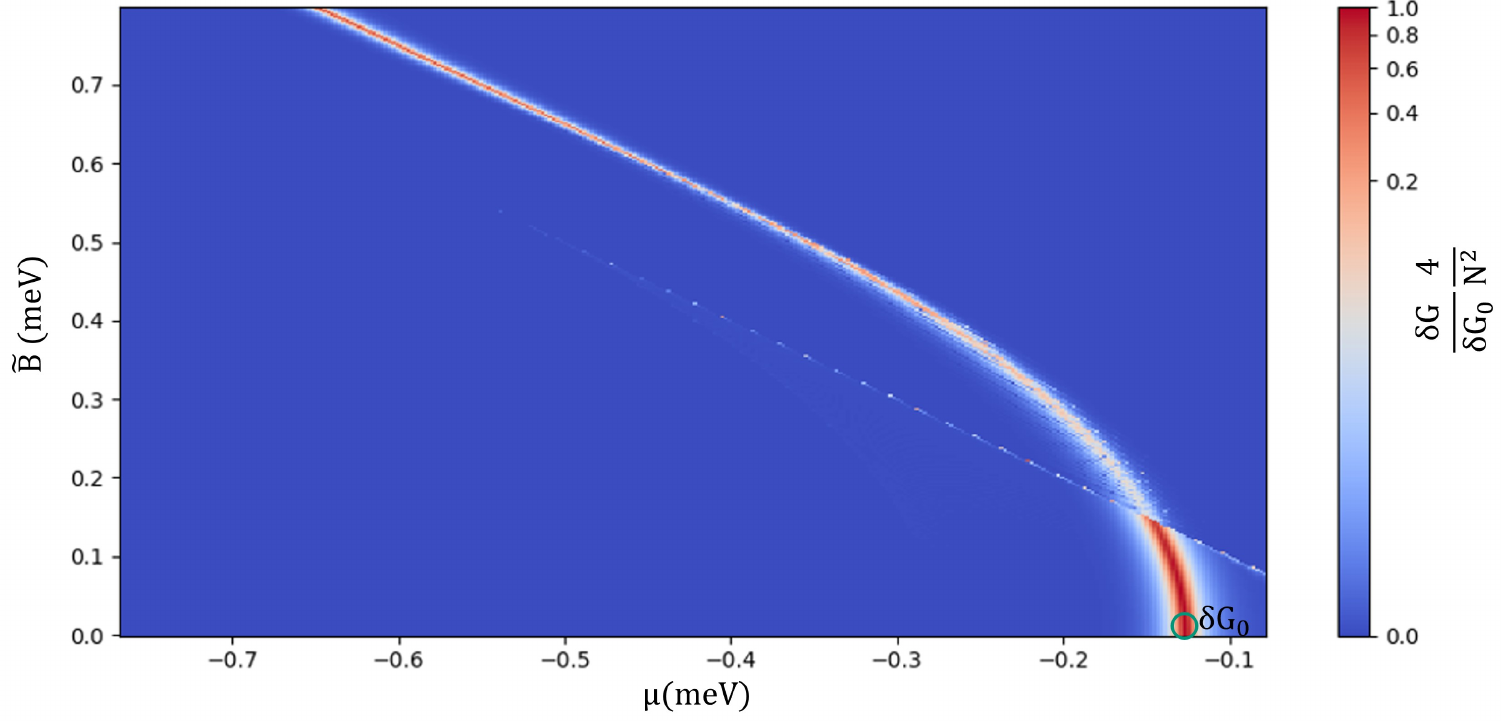}
\caption{\label{fig:SIdeltaG2D}
A 2D color map of the conductance plotted as a function of magnetic field $\tilde{B}=g\mu_B B$ and chemical potential $\mu$.
The magnitude of $\delta G$ is significantly enhanced along the red curve in the $\mu$-$\tilde{B}$ plane, which corresponds to the set of $(\mu, \tilde{B})$ values where the channel is half-open. 
}
\end{figure}

\begin{figure}[!t]
\includegraphics[width=0.8\columnwidth]{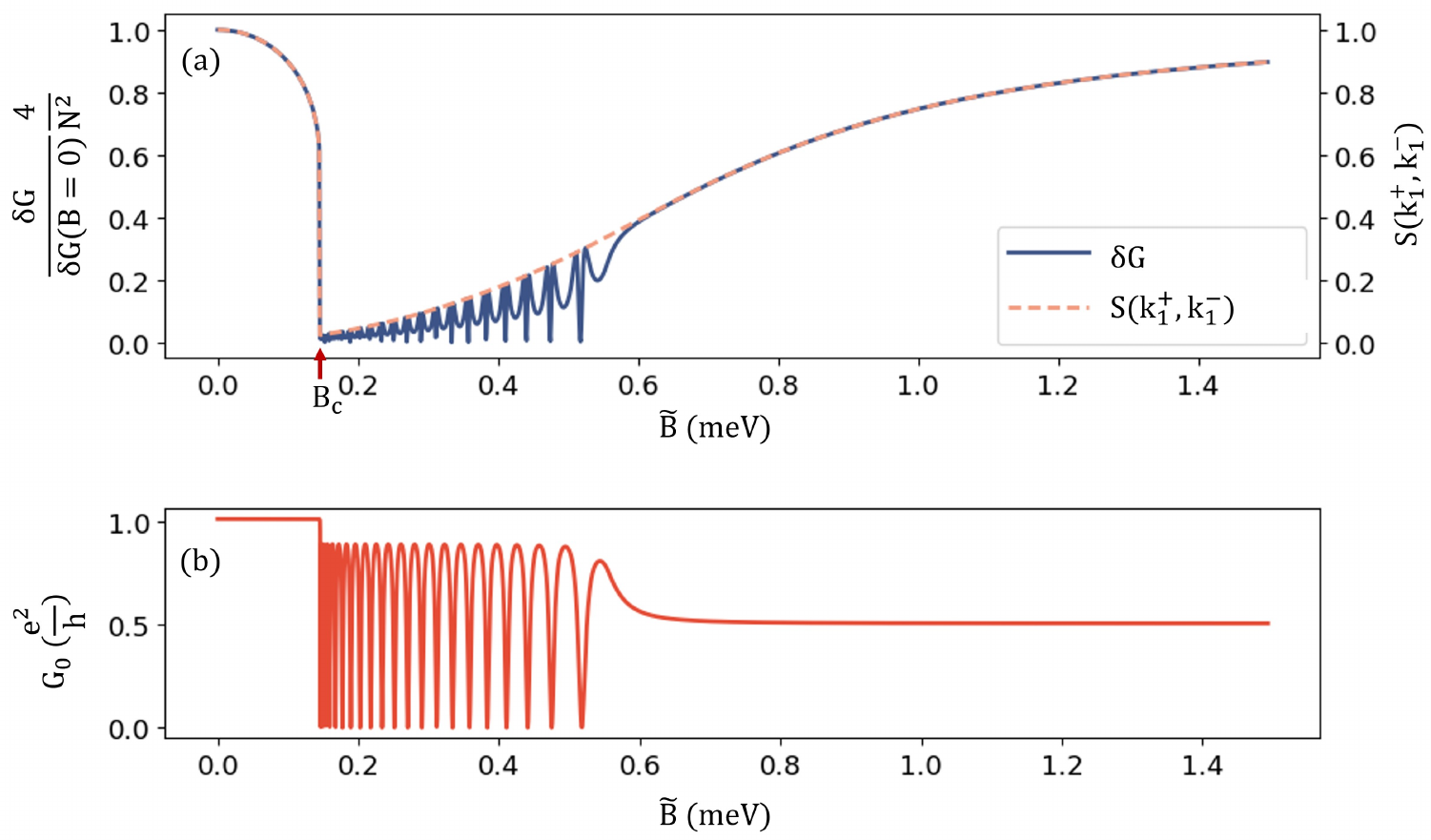}\caption{\label{fig:SIOscillatingConductance}
(a) The variation of anomalous conductance $\delta G$ along the red curve in the $\mu$-$\tilde{B}$ plane of Fig.~\ref{fig:SIdeltaG2D}. Along this curve, the channel is half-open. In this plot, when $\tilde{B}$ is increased, $\mu$ also changes to keep the channel half-open.
(b) The variation of noninteracting conductance $G_0$ along the same curve in the $\mu$-$\tilde{B}$ plane of Fig.~\ref{fig:SIdeltaG2D}.
}
\end{figure}

For a smooth barrier, such as the P\"{o}schl-Teller potential in Eq.~(\ref{eq:Vx}) of the main text, the corresponding scattering matrix in Eq.~(\ref{SI:Smatrix}) can be computed numerically. Substituting its matrix elements into Eqs.~(\ref{SI:dr11})--(\ref{SI:dr22}) yields the corrections to reflection matrix elements. These corrections are then used to determine the transmission coefficient correction in Eq.~(\ref{SIeq:deltT}) and the conductance correction $\delta G$ in Eq.~(\ref{SIeq:GT}). The resulting $\delta G$ as a function of the chemical potential $\mu$ and the magnetic field $\tilde{B}$ is plotted in Fig.~\ref{fig:SIdeltaG2D}. As shown, the magnitude of $\delta G$ is significantly enhanced along a specific curve in the $\mu$-$\tilde{B}$ plane, which corresponds to the set of $(\mu, \tilde{B})$ values where the channel is half-open. Fig.~\ref{fig:SIOscillatingConductance}(a) (identical to Fig.~\ref{fig:FullB} in the main text)  displays the variation of $\delta G$ along this curve.
Fig.~\ref{fig:SIOscillatingConductance}(b) shows the corresponding variation of the noninteracting conductance, $G_0$, characterized by a series of oscillations for $\tilde{B} > \tilde{B}_c$, the origin of which has been explained in Section \ref{SIsec:G0}. Fig.~\ref{fig:SIOscillatingConductance}(b) thereby demonstrates that the oscillations in $\delta G$ for $\tilde{B} > \tilde{B}_c$ in Fig.~\ref{fig:SIOscillatingConductance}(a) originate from the oscillations in the noninteracting reflectivity. 

\subsection{Analytical results for anomalous conductance under weak and strong magnetic fields}

When the magnetic field is weak, scattering between $k_{1}^{+}$ and $k_{2}^{-}$ (as well as $k_{2}^{+}$ and $k_{1}^{-}$) in Fig.~\ref{fig:SIdispersion} is weak; consequently, $r^{L,R}_{12,21}$ and $t^{L,R}_{12,21}$ are small. Together with the properties $r_{11}^L=r_{22}^L$, $r_{11}^R=r_{22}^R$, $\Delta k_{11}=\Delta k_{22}$, and $\left|S_{11}\right|^{2}=\left|S_{22}\right|^{2}$, the corrections to the reflection matrix elements can be further simplified to
\begin{align}
\delta r^L_{11}=&2\frac{V_0}{4\pi \hbar v_{k}}
\ln\left(\frac{1}{\left|k-k_F\right|L}\right)\left|S_{11}\right|^{2}  
\left[r_{11}^{L}\left(1-r_{11}^{L}r_{11}^{L*}\right)-r_{11}^{R*}t_{11}^{L}t_{11}^{R}
\right]  \label{SI:d2r11v1} \\
=&-\frac{V_0}{\pi\hbar v_{k}}
\ln\left(\frac{1}{\left|k-k_F\right|L}\right)\left|S_{11}\right|^{2}  
r_{11}^{L}\mathcal{T}_0,
\label{SI:d2r11}
\end{align}
\begin{align}
\delta r^L_{22}=&2\frac{V_0}{4\pi \hbar v_{k}}
\ln\left(\frac{1}{\left|k-k_{F}\right|L}\right)\left|S_{22}\right|^{2}  
\left[r_{22}^{L}\left(1-r_{22}^{L}r_{22}^{L*}\right)-r_{22}^{R*}t_{22}^{L}t_{22}^{R}
\right] \label{SI:d2r22v1}  \\
=&-\frac{V_0}{\pi\hbar v_{k}}
\ln\left(\frac{1}{\left|k-k_{F}\right|L}\right)\left|S_{22}\right|^{2}  
r_{22}^{L}\mathcal{T}_0,
\label{SI:d2r22}
\end{align}
\begin{align}
\delta r^L_{21}\approx0,
\end{align}
\begin{align}
\delta r^L_{12}\approx 0.
\end{align}
Here $k=\left(k_{1}^{+}-k_{2}^{+}\right)/2$, and the factor of $2$ has been absorbed into the length $L$.
The unitary property of the scattering matrix has been used in going from Eq.~(\ref{SI:d2r11v1}) to Eq.~(\ref{SI:d2r11}), as well as from Eq.~(\ref{SI:d2r22v1}) to Eq.~(\ref{SI:d2r22}). Here, $\mathcal{T}_0$ is given by $\mathcal{T}_0=1-\left|r^L_{11}\right|^2=1-\left|r^L_{22}\right|^2$.

Using Eqs.~(\ref{SI:d2r11}) and (\ref{SI:d2r22}), we find
\begin{align}
\text{Re}(r_{11}^{L*}\delta r^L_{11})=\text{Re}r_{22}^{L*}\delta r^L_{22})
=\frac{V_0}{\pi\hbar v_{k}}
\ln\left(\frac{1}{\left|k-k_{F}\right|L}\right)\left|S_{11}\right|^{2} \mathcal{T}_0\left(1-\mathcal{T}_0\right).
\end{align}
The resulting contribution to the anomalous transmission at $k$ is
\begin{align}
\delta \mathcal{T}_k=-4\frac{V_0}{\pi\hbar v_{k}}
\ln\left(\frac{1}{\left|k-k_{F}\right|L}\right)\left|S_{11}\right|^{2}  \mathcal{T}_0\left(1-\mathcal{T}_0\right).
\end{align}
The anomalous conductance at finite temperature can be obtained through Eq.~(\ref{SIeq:GT}).
At low temperature, $\ln\left(\frac{1}{\left|k-k_{F}\right|L}\right)\left(-\frac{\partial f_{\epsilon}}{\partial \epsilon}\right)$ is a sharp peak at the Fermi energy, and $\delta G$ can be further simplified to
\begin{align}
\delta G =- 4\frac{e^2}{h}\frac{V_0}{\pi\hbar v_{k_F}} \kappa_T \left|S_{11}\right|^{2}  \mathcal{T}_0\left(1-\mathcal{T}_0\right).\label{dGtwochannel}
\end{align}
At temperature $T$, 
\begin{align}
\kappa_T=\int 
\ln\left(\frac{1}{\left|k\left(\epsilon\right)-k_{F}\right|L}\right) \left( -\frac{\partial f_{\epsilon}}{\partial \epsilon} \right) d\epsilon.
\end{align}
As can be seen from Eq.~(\ref{dGtwochannel}), the anomalous conductance at small magnetic fields is governed by the spin overlap $\left|S_{11}\right|^2$.


When the magnetic field becomes large, the number of Fermi points will be reduced from $4$ to $2$, as shown in Fig.~\ref{fig:Dispersion}(c) in the main text. The scattering matrix~(\ref{SI:Smatrix}) will be reduced to $S=\left(\begin{array}{cc}
r_{21}^{L} & t_{22}^{R}\\
t_{11}^{L} & r_{12}^{R}
\end{array}\right)$.
In Fig.~\ref{fig:SIdispersion}, if the magnetic field is large enough, the remaining two Fermi points will be $k_{1}^{+}$ and $k_{2}^{-}$, and the wave function at $x\ll \ell(\epsilon)$ is then
\begin{align}
\Psi_{k}^{L}\left(x\right)=e^{ik_{1}^{+}x}\left|\chi_{1}^{+}\right\rangle
+\left(r_{21}^{L}+\delta r^L_{21}\right)e^{ik_{2}^{-}x}\left|\chi_{2}^{-}\right\rangle, 
\end{align}
where the correction to the reflection is
\begin{align}
\delta r^L_{21}=&\frac{V_0}{4\pi \hbar v_{k}}
\ln\left(\frac{1}{\left|k-k_{F}\right|L}\right)
\left|S_{12}\right|^{2}\left(r_{21}^{L}-r_{21}^{L}r_{21}^{L*}r_{21}^{L}-r_{12}^{R*}t_{11}^{L}t_{22}^{R}\right) \nonumber \\
=&\frac{V_0}{2\pi \hbar v_{k}}
\ln\left(\frac{1}{\left|k-k_{F}\right|L}\right)
\left|S_{12}\right|^{2}r_{21}^{L}\mathcal{T}_0.
\end{align}
Here $\mathcal{T}_0=1-\left|r^L_{21}\right|^2$, $k=\left(k_{1}^{+}-k_{2}^{-}\right)/2$. 
Following the same procedure to obtain Eq.~(\ref{dGtwochannel}), one can find the anomalous conductance in this case at low temperature:
\begin{align}
\delta G =- \frac{e^2}{h}\frac{V_0}{\pi\hbar v_{k_F}} \kappa_{T} \left|S_{12}\right|^{2}  \mathcal{T}_0\left(1-\mathcal{T}_0\right).
\end{align}
As seen from this equation, at large magnetic fields, the anomalous conductance is governed by the spin overlap $\left|S_{12}\right|^2$.

\end{document}